\documentclass[11pt,a4paper]{article}
\usepackage{hyperref}
\usepackage{amsmath} 
\usepackage{amsmath,amssymb} 
\usepackage{graphicx} 
\usepackage[left=2.9cm,right=2.9cm]{geometry}
\usepackage{subfig}
\usepackage[dvipsnames]{xcolor}
\usepackage[maxnames=10,minnames=1,backend=biber,style=numeric-comp,sorting=none]{biblatex}
\usepackage{comment}
\begin{document}

\title{Thermal Radiation from an Analytic Hydrodynamic Model with Hadronic and QGP Sources in Heavy-Ion Collisions}

\author{
Gábor László Kasza\thanks{Email: \texttt{kasza.gabor@wigner.hun-ren.hu}}\\
HUN-REN Wigner RCP, H-1525 Budapest 114, POB 49, Hungary\\
MATE Institute of Technology KRC, H-3200 Gy{\"o}ngy{\"o}s, Hungary
}

\maketitle

\begin{abstract}%
In high-energy heavy-ion collisions, a nearly perfect fluid is formed, known as the strongly coupled quark-gluon plasma (QGP). After a short thermalization period, the evolution of this medium can be described by the equations of relativistic hydrodynamics. As the system expands and cools, the QGP undergoes a transition into hadronic matter, marking the onset of quark confinement. Direct photons offer insights into an essential stage of evolution, spanning from the onset of thermalization to the suppression of thermal photon production, which occurs within the hadronic phase. This paper builds upon and extends a previously published solution of relativistic hydrodynamics, incorporating an equation of state that falls within the same class as that predicted by lattice QCD. Based on this solution, a completely analytic model is constructed to describe thermal photon production, accounting for the quark-hadron transition. The model is tested against PHENIX measurements of non-prompt direct photon spectra in Au+Au collisions at $\sqrt{s_{NN}} = 200$ GeV. Good agreement is observed between the model predictions and the experimental data, enabling the investigation of the centrality dependence of the initial temperature. These results provide a benchmark for future theoretical and experimental studies of thermal radiation in heavy-ion collisions.
\end{abstract}

\section{Introduction}
One of the central objectives of high-energy heavy-ion physics is to investigate the state of matter that existed within fractions of a millisecond after the Big Bang. During this primordial epoch, the Universe was characterized by extreme temperatures and pressures, under which ordinary matter could not exist. Instead, the Cosmos was permeated by an exotic phase of matter composed of deconfined quarks and gluons — a state referred to as the strongly coupled quark-gluon plasma (QGP). In this medium, the quark-quark scattering cross section becomes divergent, and the mean free path approaches zero, indicating that the QGP behaves not as an ideal gas but rather as a nearly perfect fluid of quarks~\cite{Lacey:2006bc}.

Hydrodynamic equations, being scale-invariant, are applicable across a vast range of length scales — from the smallest distances accessible in experiments to cosmological scales. Consequently, relativistic hydrodynamics provides a powerful framework for modeling the spacetime evolution of the hot and dense matter produced in high-energy heavy-ion collisions, where femtometre-scale fireballs are created and evolve according to these principles.

However, hadronic observables offer limited insight into the dynamic properties of the QGP itself, as the measured hadron spectra primarily reflect the conditions at the time of kinetic freeze-out. To probe the earlier stages of the system’s evolution or to gain information about its equation of state, it is advantageous to consider observables based on other types of particles. Direct photons serve as particularly valuable messengers in this context.

In heavy-ion collisions, real photons are emitted from multiple sources. Broadly, one distinguishes between decay photons and direct photons. Decay photons originate from the electromagnetic decays of long-lived hadronic final states. In contrast, direct photons are produced at any stage of the collision process and include both prompt and thermal contributions~\cite{David:2019wpt}. The prompt component typically arises from initial hard parton-parton scatterings, and also includes contributions from pre-equilibrium processes~\cite{Shuryak:1992bt,Tuchin:2010gx,McLerran:2014hza,Zakharov:2016kte,Ayala:2017vex,Berges:2017eom,David:2019wpt,Gale:2021emg,Monnai:2022hfs,Garcia-Montero:2023ubi,Garcia-Montero:2024msw}. Thermal photons, on the other hand, are emitted from a locally thermalized, collectively expanding medium. At each spacetime point of emission, the medium can be characterized by a well-defined temperature and flow profile. These photons constitute genuine electromagnetic radiation from the hot and dense matter and are not decay products. Depending on the phase of the medium at the time of their emission, they are classified as originating either from the QGP phase or from the hadronic phase.

Due to their negligible interaction probability with the strongly interacting medium, direct photons can traverse the system largely unaltered. As a result, they retain information about the local conditions at their point of emission, such as temperature and collective velocity fields, making them excellent probes of the early evolution of the medium.

This work focuses on the description of the thermal photon component using a newly developed hydrodynamical model. The model predictions are compared with recent measurements of direct photon transverse momentum $p_{\rm T}$ spectra in $\sqrt{s_{\rm NN}}=200$ GeV $Au+Au$ collisions as recorded by the PHENIX experiment~\cite{PHENIX:2022rsx}. At high $p_{\rm T}$, the direct photon spectrum is dominated by prompt contributions, while at low $p_{\rm T}$, thermal radiation becomes the leading source (although unconventional photon production mechanisms may also play a role; see refs.~\cite{Berges:2017eom,Heffernan:2014mla,vanHees:2014ida,Muller:2013ila,Basar:2012bp,Basar:2014swa}). As such, hydrodynamical models are expected to describe only the low-$p_{\rm T}$ region of the spectrum, or more precisely, spectra from which the hard scattering contribution has been subtracted. Fortunately, PHENIX has performed this subtraction, allowing the present analysis to compare the hydrodynamic model not to the low-$p_{\rm T}$ direct photon yield, but specifically to the so-called non-prompt component.

The new hydrodynamical model introduced in this manuscript is constructed entirely within an analytical framework. It builds, on the one hand, on a previously published exact and analytic solution of relativistic perfect fluid hydrodynamics in 1+1 (temporal and longitudinal) dimensions, which features a locally accelerating velocity field, along with a novel generalization thereof; and, on the other hand, on the analytic computation of the transverse momentum spectrum of thermal radiation.

Naturally, the question arises why the $p_{\rm T}$ spectra of thermal photons are analyzed within a 1+1 dimensional model that lacks transverse dynamics. The reason is that the 
$p_{\rm T}$ spectrum of thermal photons originates directly from the temperature distribution of the medium, and since photons essentially do not interact after their production, their spectrum is highly sensitive to the temperature evolution of the fireball. While transverse flow does influence the $p_{\rm T}$ distribution of direct photons, in the early-time regime relevant for determining the initial temperature, a significant contribution comes from the temperature of the medium and its time evolution. As shown in Fig.~2 of ref.~\cite{Shen:2013vja}, the impact of radial flow is clearly visible across the entire photon spectrum, although it is less pronounced for the high-temperature components than for the low-temperature ones. Therefore, although the present discussion is based on a 1+1 dimensional framework that does not include transverse flow, it yields valuable qualitative information on the initial temperature and the space-time evolution of the medium, while maintaining relatively simple analytical expressions. Accordingly, analyzing thermal photon $p_{\rm T}$ spectra within this 1+1 dimensional framework remains suitable for extracting qualitative information on the initial temperature of the medium.

Compared to earlier analytic hydrodynamic models that aimed to describe direct photon spectra or its thermal component~\cite{Csanad:2011jq,Csanad:2011jr,Kasza:2023rpx,Lokos:2024yjm}, the novelty of the present approach lies in its treatment of the medium as evolving through multiple distinct phases. Accordingly, the model developed here describes the thermal photon spectrum as the sum of hadronic and QGP components. Previously, a single-component analytic model, considered as a predecessor to the present work, was published in ref.~\cite{Kasza:2023rpx}. While that model was able to describe the PHENIX data with a high confidence level, it predicted the initial temperature only with large uncertainties and tended to overestimate its value.

One of the key motivations for developing a new analytic model arises from the observation that experimental data cannot be adequately described by a simple Boltzmann spectrum with a constant inverse slope parameter~\cite{PHENIX:2022rsx,Esha:2022tvk,Orosz:2025cey}. As a consequence, the inverse slope of the measured $p_{\rm T}$ spectra does not directly represent any well-defined temperature of the system~\cite{Shen:2013vja, Linnyk:2013hta, Paquet:2017wji, David:2019wpt}. Therefore, the determination of the initial temperature of the QGP requires more sophisticated models, whose development poses a considerable challenge if one aims to stay entirely within an analytic framework.

As outlined above, the new model separates the thermal radiation into QGP and hadronic contributions. This decomposition is particularly advantageous because, similarly to hadrons, thermal photons carry imprints of radial flow, being emitted from a collectively expanding medium in motion. Due to the Doppler effect, the motion of the emission source leads to a characteristic blue-shifting, or "blow-up," of the $p_{\rm T}$ spectrum~\cite{David:2019wpt,Monnai:2022hfs,Massen:2024pnj}. While this effect is present for all thermal sources, it is more pronounced for the lower-temperature contributions and therefore impacts the hadronic component more strongly than the early-time QGP contribution. Since the present model is based on a 1+1 dimensional solution, incorporating only longitudinal dynamics, the computed spectra do not include the effects of radial flow. Consequently, the separation of the hadronic component (which is more affected by radial flow) from the QGP contribution is especially beneficial for the determination of the initial temperature, as the relevant information is encoded in the thermal photon yield of the QGP.

The question of where, in the transverse momentum spectrum of thermal radiation, the hadronic contribution begins to outshine that of the QGP remains an open and actively investigated topic~\cite{David:2019wpt}. Answering this question poses a significant challenge, as the direct photon spectrum is a superposition of multiple sources that are experimentally difficult to separate. Early estimates assumed that the dominant source of thermal radiation in relativistic heavy-ion collisions was the QGP. This assumption was later challenged in ref.~\cite{Kapusta:1991qp}, which emphasized the potential significance of hadrons as sources of thermal photons. Around the year 2000, the direct photon results from the WA98 experiment in Pb+Pb collisions~\cite{WA98:2000vxl}, together with advancements in the quark–hadron duality hypothesis and in the modeling of the fireball’s space-time evolution using Hwa-Bjorken hydrodynamics~\cite{Hwa:1974gn, Bjorken:1982qr}, contributed to the emergence of more refined theoretical calculations. For example, refs.~\cite{Steffen:2001pv,Linnyk:2015tha} predicted that in RHIC collisions, the QGP contribution becomes dominant above $p_{\rm T}\approx 3 $ GeV, while below this threshold, radiation from hadronic matter is expected to prevail. However, for the same question in the case of LHC collisions, refs.~\cite{Steffen:2001pv} and~\cite{Linnyk:2015tha} provide differing predictions, although ref.~\cite{Linnyk:2015tha} does not allow for a clear conclusion due to the large uncertainties associated with the yield of QGP. Slightly differing conclusions have been reached in ref.~\cite{Srivastava:1999ekv}, where it was argued that even below 1 GeV, quark matter remains the dominant source of radiation. In a study from 2003, Turbide, Rapp, and Gale reported that in $\sqrt{s_{\rm NN}} = 200$ GeV $Au+Au$ and $\sqrt{s_{\rm NN}} = 5.5$ TeV $Pb+Pb$ collisions, the hadronic source begins to dominate the thermal radiation only below $p_{\rm T}\approx 1$ GeV~\cite{Turbide:2003si}.

Due to the absence of transverse dynamics in the present model, elliptic flow - quantified by the coefficient $v_2$ - cannot be addressed within this framework. As such, the model is not suitable for investigating the so-called "direct photon puzzle" - a term used to describe the experimental finding that the elliptic flow coefficient of direct photons is of similar magnitude to that of hadrons~\cite{David:2019wpt,Monnai:2022hfs,Gale:2018ofa}. This remains a striking observation, as it is not yet fully understood how photons emitted during the early, hotter stages of the collision — when collective flow is expected to be weak — can exhibit a $v_2$ comparable to that of hadrons, which are produced in later, cooler phases. While sophisticated numerical models exist that successfully describe the photon spectrum, they tend to underestimate the observed elliptic flow~\cite{Linnyk:2015tha,Gale:2014dfa,Chatterjee:2017akg}. A closely related recent work is presented in ref.~\cite{Lokos:2024yjm}, reporting promising findings in this context: in collaboration with my co-author, we employed a 1+3 dimensional semi-analytic hydrodynamic model to simultaneously describe both the direct photon spectrum and the elliptic flow measured by the ALICE experiment in $\sqrt{s_{\rm NN}}=2.76$ TeV Pb+Pb collisions~\cite{ALICE:2015xmh,ALICE:2018dti}.

The manuscript is structured as follows. In Section~\ref{sec:2}, I present new solutions of relativistic perfect fluid hydrodynamics, one tailored for describing hadronic fireballs and the other for QGP-phase fireballs. These solutions can be regarded as generalizations of the one introduced in ref.~\cite{Csorgo:2018pxh}, in the sense that they impose less restrictive conditions on the form of the equation of state. Section~\ref{sec:3} details how the two-component transverse momentum ($p_{\rm T}$) spectrum of thermal radiation can be derived from the new solutions, including a discussion of the analytical approximations applied. In Section~\ref{sec:4}, the two-component spectrum is compared — using two distinct approaches — to the non-prompt component of the direct photon spectrum measured by the PHENIX Collaboration in $Au+Au$ collisions at $\sqrt{s_{\rm NN}} = 200$ GeV, across four different centrality classes~\cite{PHENIX:2022rsx}. Particular attention is given to the assumptions used to constrain or fix the values of certain fit parameters. In Section~\ref{sec:5}, I examine whether the new model introduced in this manuscript is capable of describing hadronic observables using a set of fit parameters that are consistent with those obtained from the fit of non-prompt direct photon spectra. In Section~\ref{sec:6}, I investigate the equation of state predicted by the new model when its parameters are constrained based on experimental data and lattice QCD simulations. In Section~\ref{sec:7}, I compare the results obtained in Section~\ref{sec:4} primarily, but not exclusively with analytic hydrodynamic models. The final section summarizes the manuscript and offers a brief outlook on possible future developments of the proposed model.

\section{New Solutions of Relativistic Hydrodynamics}\label{sec:2}
In this section, I present new solutions of relativistic hydrodynamics, which are generalizations of the 1+1 dimensional perfect fluid solution reported in ref.~\cite{Csorgo:2018pxh}. This earlier solution was found by assuming an equation of state in which the proportionality factor $\kappa$ between the energy density $\varepsilon$ and the pressure $p$ is constant. In this paper, I consider $\kappa$ as a temperature-dependent function and adopt an equation of state that belongs to the same class as that found in lattice QCD calculations~\cite{Borsanyi:2010cj}. 

It should be emphasized, that while numerical hydrodynamic simulations commonly employ a temperature-dependent ratio $\kappa = \varepsilon/p$, most analytic hydrodynamic solutions widely used in heavy-ion physics assume a constant $\kappa$. Notable examples include the Hwa–Bjorken~\cite{Hwa:1974gn,Bjorken:1982qr}, the Landau–Khalatnikov~\cite{Landau:1953gs,Khalatnikov:1954}, the Gubser~\cite{Gubser:2010ui,Gubser:2010ze}, the \linebreak Białas–Janik–Peschanski~\cite{Bialas:2007iu}, and the Csörgő–Grassi–Hama–Kodama~\cite{Csorgo:2003rt} solutions. From this perspective, deriving a closed (inhomogeneous) analytic solution that incorporates a temperature-dependent $\kappa(T)$ constitutes a significant advancement.

I first summarize the system of conservation equations of hydrodynamics in a form that takes into account the assumed velocity field and the more general form of the equation of state, and then I present two distinct solutions to these equations, aimed at describing the evolution of the thermalized fireball during the QGP and hadron phases.

\subsection{Conservation Equations and the Equation of State}
For perfect fluids, relativistic hydrodynamics describes the local conservation laws of energy, momentum, particle number and entropy:
\begin{align}
    \partial_{\nu}T^{\mu\nu} &= 0, \label{eq:energy-momentum-cons} \\
    \partial_{\mu}\left(n u^{\mu}\right) &= 0,\\
    \partial_{\mu}\left(\sigma u^{\mu}\right) &= 0,
\end{align}
where $T^{\mu \nu}$ is the energy-momentum four-tensor of perfect fluids, $n$ is the particle density, the entropy density is denoted by $\sigma$ and $u^{\mu}$ stands for the four-vector of the velocity field. In order to render the hydrodynamic system of differential equations solvable, we supplement it with an equation of state defined by the following expressions:
\begin{align}
    \varepsilon &= \kappa(T)p,\label{eq:eos_eps}\\ 
    \mu_B &= 0, \label{eq:eos_mu}
\end{align}
where $\mu_B$ is the bariochemical potential. This equation of state remains thermodynamically consistent for an arbitrary functional form of $\kappa(T)$, as can be verified using the free energy density $f(T)$. At vanishing bariochemical potential, the fundamental thermodynamic relation reads
\begin{equation}
\varepsilon + p = T\sigma,
\end{equation}
which leads to the following differential equation for the temperature dependence of the pressure:
\begin{equation}
p(T) = \frac{T}{1+\kappa(T)}\frac{\partial p}{\partial T}.
\end{equation}
The solution of this equation is
\begin{equation}
p(T) = p_{\rm i}\frac{T}{T_{\rm i}}\exp\left(\int\limits_{T_{\rm i}}^{T}\frac{\kappa(T')}{T'}dT'\right),
\end{equation}
where $p_{\rm i}$ and $T_{\rm i}$ denote integration constants. One readily verifies that the following relations between the free energy density $f$, the pressure $p$, and the energy density $\varepsilon$ hold:
\begin{align}
p &= -f,\\
\varepsilon &= f - T\frac{\partial f}{\partial T},
\end{align}
thereby ensuring the consistency of the equation of state defined in eqs.~\eqref{eq:eos_eps} and~\eqref{eq:eos_mu}.

Based on the assumption regarding the equation of state, the energy conservation equation and the relativistic Euler equation — corresponding to the projections of Eq.~\eqref{eq:energy-momentum-cons} parallel and perpendicular to $u^{\mu}$, respectively — take the following form when expressed in terms of temperature:
\begin{align}
    \left[\kappa(T)+\frac{T}{1+\kappa(T)}\frac{\partial\kappa(T)}{\partial T}\right]u^{\mu}\partial_{\mu}T + T\partial_{\mu}u^{\mu}&=0,\label{eq:energy-cons} \\ 
    T u^{\mu} \partial_{\mu} u^{\nu} - \left(g^{\mu\nu} - u^{\mu}u^{\nu}\right)\partial_{\mu} T &= 0.
\end{align}
When $\mu_B=0$, a trivial solution of the particle number continuity equation is $n=0$, since $n$ denotes the net particle density, i.e., the difference between particle and antiparticle densities.

It can be observed that among the conservation equations, only the energy conservation (eq.~\eqref{eq:energy-cons}) depends on the parameter $\kappa$, which in this work is allowed to vary with temperature. Notably, when $\kappa$ is taken to be constant, the resulting equation reduces to the form, which was solved analytically in ref.~\cite{Csorgo:2018pxh} by using a specific velocity field. Consequently, if $\kappa = \kappa(T) \neq \textnormal{const.}$ and the following equation holds:
\begin{equation}
    \kappa(T)+\frac{T}{1+\kappa(T)}\frac{\partial\kappa(T)}{\partial T} = C =\textnormal{const.}, \label{eq:kappa-diff-eq}
\end{equation}
then, a solution similar to that given in ref.~\cite{Csorgo:2018pxh} can be derived, except that the equations describing the new solution will contain the constant from the right-hand side of eq.~\eqref{eq:kappa-diff-eq} instead of a constant $\kappa$. However, this approach imposes a stringent constraint on the form of $\kappa(T)$ as expressed in eq.~\eqref{eq:kappa-diff-eq}. Fortunately, this differential equation can be easily integrated, yielding the following result:
\begin{equation}\label{eq:eos_def}
    \kappa(T) = \frac{C\left(\frac{T}{T_0}\right)^{1+C}-\frac{C-\kappa_0}{\kappa_0+1}}{\left(\frac{T}{T_0}\right)^{1+C} + \frac{C-\kappa_0}{\kappa_0+1}},
\end{equation}
where $T_0$ is the lower limit of integration and $\kappa_0$ represents $\kappa(T_0)$. In this equation, the values of $C$, $T_0$, and $\kappa_0$ are all arbitrary parameters, however, the analytical structure of eq.~\eqref{eq:eos_def} suggests that the parameter $C$ corresponds to the high-temperature limit of the function $\kappa(T)$. The parameters $C$, $T_0$, and $\kappa_0$ should be adjusted to be consistent with the characteristics of the hadronic matter.

A similar procedure can be applied to the phase of quark matter; however, assigning $C=3$ in this regime is physically well motivated, as the equation of state tends toward the conformal limit $\varepsilon = 3 p$ at asymptotically high temperatures. Naturally, the parameters $\kappa_0$ and $T_0$ in this case must be chosen to be consistent with the thermodynamic behavior of the QGP.

Based on these results, we can construct a simple model for the equation of state of the thermalized medium that incorporates the quark-hadron transition. This is achieved by introducing an average transition temperature, $T_{\rm tr}$, at which the equations of state for the hadronic and QGP phases are seamlessly combined using Heaviside step functions\footnote{More sophisticated approaches can certainly be employed (for example, introducing temperature-dependent weighting factors); however, in most cases, this would require spatial homogeneity of the temperature. A solution with spatially homogeneous temperature, however, cannot describe the direct photon spectra. A more detailed discussion on this topic can be found in ref.~\cite{Lokos:2024yjm}.}$^{,}$\footnote{Since eq.~\eqref{eq:kappa-diff-eq} is a nonlinear differential equation, $\kappa_{\rm tot}$ is not itself a solution to it; only $\kappa_{\rm q}$ and $\kappa_{\rm h}$ can be regarded as valid solutions. Accordingly, the introduction of $\kappa_{\rm tot}$ is not intended to yield a hydrodynamic solution that is able to describe the temperature regimes of both the QGP and hadronic matter (i.e., it is not the result of combining different exact solutions). Rather, it provides a composite functional form that formally encodes which of the two functions—$\kappa_{\rm q}$ or $\kappa_{\rm h}$—applies in each temperature range.}:
\begin{equation}\label{eq:total-eos}
    \kappa_{\rm tot}(T) = \Theta\left(T-T_{\rm tr}\right)\kappa_{\rm q}\left(T\right) + \Theta\left(T_{\rm tr}-T\right)\kappa_{\rm h}(T),
\end{equation}
where
\begin{align}
    \kappa_{\rm h}(T) &= \frac{C_{\rm h}\left(\frac{T}{T_{\rm fo}}\right)^{1+C_{\rm h}}-\frac{C_{\rm h}-\kappa_{\rm fo}}{\kappa_{\rm fo}+1}}{\left(\frac{T}{T_{\rm fo}}\right)^{1+C_{\rm h}} + \frac{C_{\rm h}-\kappa_{\rm fo}}{\kappa_{\rm fo}+1}},\label{eq:eos_def_hadron}\\
    \kappa_{\rm q}(T) &= \frac{C_{\rm q}\left(\frac{T}{T_{\rm tr}}\right)^{1+C_{\rm q}}-\frac{C_{\rm q}-\kappa_{\rm tr}}{\kappa_{\rm tr}+1}}{\left(\frac{T}{T_{\rm tr}}\right)^{1+C_{\rm q}} + \frac{C_{\rm q}-\kappa_{\rm tr}}{\kappa_{\rm tr}+1}},\label{eq:eos_def_quark}
\end{align}
with $\kappa_{\rm fo}=\kappa_{\rm h}(T_{\rm fo})$ and $\kappa_{\rm tr}=\kappa_{\rm q}(T_{\rm tr})$. I define the matching condition between these two functions by imposing the equality $\kappa_{\rm q}(T_{\rm tr}) = \kappa_{\rm h}(T_{\rm tr})$. The parameter $T_{\rm fo}$ of the function $\kappa_{\rm h}$ can be interpreted as the kinetic freeze-out temperature of hadrons. 

At first glance, the functions $\kappa_{\rm h}$ and $\kappa_{\rm q}$ appear to exhibit very similar characteristics. However, it is crucial to highlight a key difference that distinguishes their monotonic behavior within the physically relevant temperature ranges. Specifically, in the phase of quark matter, $\kappa_{\rm q}$ must approach the limit $C_{\rm q}$ from above, whereas on the hadronic side, $\kappa_{\rm h}$ must approach the limit $C_{\rm h}$ from below~\cite{Csorgo:2016ypf,Csorgo:2018tsu}. Consequently, the signs of $C_{\rm q}-\kappa_{\rm tr}$ and $C_{\rm h}-\kappa_{\rm fo}$ are different, causing $\kappa_{\rm h}$ to increase strictly monotonically for $T>T_{\rm fo}$, while $\kappa_{\rm q}$ decreases strictly monotonically for $T>T_{\rm tr}$.

\subsection{The Solutions for the QGP and Hadronic Matter}\label{sec:solution}
Given that we are considering 1+1 dimensional solutions with dynamics confined to the longitudinal spatial direction, it is appropriate to employ Rindler coordinates, the longitudinal proper time $\tau$ and the space-time rapidity $\eta_z$:
\begin{align}
    \eta_z &= \frac{1}{2}\ln\left(\frac{t+r_z}{t-r_z}\right),\\
    \tau &= \sqrt{t^2-r_z^2},
\end{align}
where $r_z$ denotes the longitudinal spatial coordinate and $t$ represents the time coordinate. To simplify the notation in the equations that follow, we also define the fluid rapidity in terms of the longitudinal component of the velocity three-vector $v_z$:
\begin{equation}
    \Omega = \frac{1}{2}\ln\left(\frac{1+v_z}{1-v_z}\right).
\end{equation}
Following the formulation in ref.~\cite{Csorgo:2018pxh}, the fluid rapidity is assumed to be independent on the longitudinal proper time $\tau$ (i.e. $\Omega \equiv \Omega\left(\eta_z\right)$) and the locally accelerating velocity field can be written as:
\begin{equation}
    u^{\mu} = \left(\cosh\left(\Omega\right),\sinh\left(\Omega\right)\right).
\end{equation}
In the search for the new solution, I assumed the equation of state given by eq.~\eqref{eq:total-eos}. As a consequence of this assumption, two distinct solutions emerge, connected at $T_{\rm tr}$: one defined in the hadronic phase, i.e., in the temperature range $T_{\rm fo}<T<T_{\rm tr}$, which corresponds to the equation of state $\varepsilon = \kappa_{\rm h}(T) p$; and another one describing the quark matter phase, valid in the domain $T > T_{\rm tr}$, characterized by the equation of state $\varepsilon = \kappa_{\rm q}(T)p$. Since both $\kappa_{\rm h}$ and $\kappa_{\rm q}$ satisfy eq.~\eqref{eq:kappa-diff-eq}, the solutions in both the hadronic and QGP phases can be formulated analogously to the solution presented in ref.~\cite{Csorgo:2018pxh}. The only modification required is to replace the constant $\kappa$ in the solution of ref.~\cite{Csorgo:2018pxh} with the parameters $C_{\rm h}$ or $C_{\rm q}$, depending on the phase under consideration\footnote{In ref.~\cite{Csorgo:2018pxh}, the solutions for entropy density and temperature were multiplied by an arbitrary scaling function; however, for the sake of simplicity, I choose this function to be unity in the present work.} — that is, one must perform the substitution $\kappa \longleftrightarrow C_i$, where $i \in \{{\rm q},{\rm h}\}$. Following this reasoning, the solutions can be straightforwardly derived. In the hadronic phase, the equations take the following form:
\begin{align}\label{eq:hadron-solution}
    \varepsilon &= \kappa_{\rm h}(T) p,\\
    \sigma\left(\tau,\eta_z\right) &= \sigma_{\rm tr}\left(\frac{\tau_{\rm tr}}{\tau}\right)^{\lambda}\left[1+\frac{C_{\rm h}-1}{\lambda-1}\sinh^2\left(\Omega\left(\eta_z\right)-\eta_z\right)\right]^{-\frac{\lambda}{2}}, \\
    T_{\rm h}\left(\tau,\eta_z\right) &= T_{\rm tr,0} \left(\frac{\tau_{\rm tr}}{\tau}\right)^{\frac{\lambda}{C_{\rm h}}}\left[1+\frac{C_{\rm h}-1}{\lambda-1}\sinh^2\left(\Omega\left(\eta_z\right)-\eta_z\right)\right]^{-\frac{\lambda}{2C_{\rm h}}},\label{eq:temp-hadron}\\
    \Omega\left(\eta_z\right) &= \frac{\lambda}{\sqrt{\lambda-1}\sqrt{C_{\rm h}-\lambda}}\arctan\left(\sqrt{\frac{C_{\rm h}-\lambda}{\lambda-1}}\tanh\left(\Omega(\eta_z)-\eta_z\right)\right).   
\end{align}
Within the quark matter phase, the corresponding hydrodynamic solution is given by:
\begin{align}\label{eq:quark-solution}
    \varepsilon &= \kappa_{\rm q}(T) p,\\
    \sigma\left(\tau,\eta_z\right) &= \sigma_0\left(\frac{\tau_0}{\tau}\right)^{\lambda}\left[1+\frac{C_{\rm q}-1}{\lambda-1}\sinh^2\left(\Omega\left(\eta_z\right)-\eta_z\right)\right]^{-\frac{\lambda}{2}}, \\
    T_{\rm q}\left(\tau,\eta_z\right) &= T_0 \left(\frac{\tau_0}{\tau}\right)^{\frac{\lambda}{C_{\rm q}}}\left[1+\frac{C_{\rm q}-1}{\lambda-1}\sinh^2\left(\Omega\left(\eta_z\right)-\eta_z\right)\right]^{-\frac{\lambda}{2C_{\rm q}}}, \label{eq:temp-quark}\\
    \Omega\left(\eta_z\right) &= \frac{\lambda}{\sqrt{\lambda-1}\sqrt{C_{\rm q}-\lambda}}\arctan\left(\sqrt{\frac{C_{\rm q}-\lambda}{\lambda-1}}\tanh\left(\Omega(\eta_z)-\eta_z\right)\right),    
\end{align}
where the function $\Omega(\eta_z)$ can be expressed only implicitly, $\lambda$ stands for the rate of acceleration and the parameters indexed with zero ($T_0$, $\sigma_0$ and $\tau_0$) characterize the initial state in the center of the fireball. In the solution given above, which describes the hadronic phase, $\tau_{\rm tr}$ denotes the average proper time associated with the space-time location of the quark–hadron transition, and $T_{\rm tr,0}=T(\tau=\tau_{\rm tr}, \eta_z = 0)$, i.e. the temperature of the transition in the center of the fireball. This latter definition suggests that the previously introduced transition temperature $T_{\rm tr}$ depends on $\eta_z$. However, in order to make the analytical calculations feasible, I employ the approximation $T_{\rm tr}\approx T_{\rm tr,0}$. Later, in the context of calculating the direct photon spectrum, this approximation becomes equivalent to assuming that the hypersurface of the quark-hadron transition is approximately characterized by $\tau_{\rm tr}\approx \textnormal{constant}$. This simplification is justified by the fact that the dependence of temperature on $\eta_z$ is not significant within the domain of the solution, as the solution itself is only defined in a narrow region around $\eta_z = 0$.\footnote{A detailed discussion of the solution's domain of validity can be found in ref.~\cite{Csorgo:2018pxh}.} Although this approximation limits the generality of the model, it remains fully adequate for comparison with PHENIX data, which measures particles only within a narrow rapidity window.

\section{Calculating the Thermal Radiation}\label{sec:3}
In this section, I briefly outline the assumptions employed in the evaluation of the thermal photon spectrum. Thermal radiation spectra can be derived from the source function that describes the phase-space distribution of the emitting medium. It is important to emphasize that the present model differs from approaches based on relativistic kinetic theory (see, e.g., refs.~\cite{Arnold:2001ms,Dusling:2009ej,Dion:2011pp, McLerran:2014hza,Linnyk:2015rco, Paquet:2015lta,Gale:2021emg}). Instead of calculating the photon spectrum from the emission rates of microscopic interactions within the thermalized medium, I apply a collective source function for photons. 

In the present work, thermal photon emission is modeled through a four–dimensional space–time integral of a local source function. Since photons do not thermalize in the medium, we do not assume an equilibrium photon gas. Instead, we employ a Boltzmann weight, which arises from the energy distribution of the thermalized medium rather than from the photons themselves. This choice ensures the correct exponential suppression at high energies, while avoiding the misleading implication of a thermal photon ensemble. However, it is important to note that assuming a Boltzmann distribution is a highly simplified picture, particularly in the case of hadrons, as illustrated for example in the Appendix of ref.~\cite{Turbide:2003si}. I adopt this simplified assumption in order to isolate, within a single analytic framework, a lower-temperature contribution to the thermal emission, which I identify for practical purposes as the hadronic component. Naturally, in order to achieve a reliable separation between the hadronic and QGP contributions and to obtain precise quantitative predictions, more sophisticated 1+3 dimensional models based on numerical methods are required.

The locally thermalized medium itself is described using the hydrodynamic solutions introduced in Section~\ref{sec:2}. Based on these considerations, the present model describes the source of thermal radiation in terms of only two components: a low-temperature (let's call it hadronic) and a high-temperature (let's call it QGP) contribution. It does not, however, account for the individual contributions of microscopic processes, as it is formulated within a hydrodynamic framework. Nevertheless, it is informative to recall which processes are primarily responsible for thermal photon emission. In the QGP phase, the leading-order partonic channels are quark-gluon Compton scattering and quark-antiquark annihilation. In the hadronic phase, thermal photons can emerge from inelastic hadronic interactions, such as $\pi^{\pm} \rho^0 \rightarrow \pi^{\pm} \gamma$, which is the dominant channel for photon energies above 0.7 GeV in hadronic matter (particularly in the hadronic phase following heavy-ion collisions)~\cite{Kapusta:1991qp}. A more detailed discussion of the relevant microscopic reactions can be found for example in ref.~\cite{David:2019wpt}. Note that photon production mechanisms, such as gluon Compton scattering, quark–antiquark annihilation, and certain inelastic hadronic processes, exhibit a more intricate dependence on energy and temperature. A fully microscopic treatment of these processes would therefore go beyond the scope of a simple Boltzmann-type description. However, in order to remain within a tractable analytic framework, we do not consider these effects explicitly in the present study.

Admittedly, analytic hydrodynamic models, such as the one introduced in the present manuscript, are necessarily limited by their inherent simplifications and cannot reproduce the complexity and precision of numerical 2+1 or 3+1 dimensional hydrodynamic or transport models. However, the goal of the present study is different: to build a simple analytic hydrodynamic model that captures the essential features of the thermal photon spectrum, thereby offering a transparent baseline and a useful reference point for more detailed numerical approaches. The stepwise refinement of analytic models provides a bridge  between phenomenological intuition and the results of full microscopic simulations.

In light of the above considerations, I define the source function of thermal photons as follows:
\begin{equation}\label{eq:source-func}
    S\left(x^{\mu},p^{\mu}\right)d^4 x=\frac{g}{\left(2\pi \hbar\right)^3} \frac{ H(\tau)}{\tau_{\rm R}} \exp\left(-\frac{p^{\mu}u_{\mu}}{T}\right)p_{\mu}d\Sigma^{\mu},
\end{equation}
where $g$ stands for the degeneracy factor. The space-time coordinate and four-momentum of the photons are denoted by $x^{\mu}$ and $p^{\mu}$, respectively. In this source function, the properties of the thermalized medium are encoded in the temperature $T$ and the flow field $u_{\mu}$. Here, $d\Sigma^{\mu}$ denotes an infinitesimal space–time volume element comoving with the fluid, chosen to be parallel to the local four-velocity field $u^{\mu}$ of the expanding medium in accordance with the assumption of flow-aligned emission. The contraction 
$p_{\mu}d\Sigma^{\mu}$ then represents the contribution of this cell to the particle flux. In this work, thermal photon emission is treated as a volumetric process, occurring continuously throughout the space–time volume of the medium. Unlike hadrons, which are typically emitted from a freeze-out hypersurface via the Cooper–Frye formalism~\cite{Cooper:1974mv}, photons escape essentially without rescattering. Accordingly, the photon source function is integrated over the entire four-dimensional volume of the medium rather than being restricted to a freeze-out hypersurface. This approach captures the full space–time distribution of photon production and provides a clear description of volumetric photon emission.

When the medium is described by the solution presented in Section~\ref{sec:2}, the comoving volume element $d\Sigma^{\mu}$ can be expressed as follows:
\begin{equation}
    d\Sigma^{\mu} = \frac{u^{\mu}\tau d\tau d\eta_z dr_x dr_y}{\cosh\left(\Omega\left(\eta_z\right)-\eta_z\right)}.
\end{equation}
Here, it can be seen that the transverse plane is also taken into account in the calculation of the spectrum; in other words, the result will be embedded in a 1+3 dimensional spacetime. However, this comes with the assumption that the temperature is homogeneous in the transverse plane, since the solution presented in Section~\ref{sec:2} only incorporates the longitudinal dynamics.

The expression $H(\tau)/\tau_{\rm R}$ is referred to as the window function, where $\tau_{\rm R}$ is a proper-time scaling parameter introduced to ensure the correct dimensionality of the window function.  The parameter $\tau_{\rm R}$ can be interpreted as the proper-time scale over which the average particle number in a small phase-space volume reaches the value specified by the local phase-space density. The function $H(\tau)$ describes the proper time distribution corresponding to the medium’s transparency to photons. Photons are continuously produced within the medium and do not participate in the strong interaction, resulting in a mean free path that can exceed the size of the thermalized fireball. Accordingly, $H(\tau)$ represents the effective duration of thermal radiation — the proper time interval between thermalization and the fading of thermal photon emission — and is given by the expression below:
\begin{equation}
    H(\tau)=\Theta\left(\tau-\tau_{\rm 0}\right)-\Theta\left(\tau-\tau_{\rm f}\right),
\end{equation}
where $\tau_0$ denotes the initial proper time associated with the thermalization time, while $\tau_{\rm f}$ represents the average value of the proper-time domain beyond which the production of thermal photons becomes negligible. This latter event likely occurs shortly after the chemical freeze-out of hadrons, since inelastic collisions among hadrons become increasingly rare at that point, leading to a rapid decline in thermal photon production. Thus the form of the function $H(\tau)$ suggests, that the medium remains transparent to photons throughout the whole emission period, and therefore the thermal radiation does not cease due to the medium becoming opaque.

As photons escape essentially without rescattering, I omit any additional temperature-dependent polynomial prefactor that would encode detailed interaction rates. While microscopic rate calculations indeed provide such prefactors, previous studies~\cite{Csanad:2011jq, Lokos:2024yjm} have shown that the qualitative features of the photon spectrum are only weakly affected by these factors as discussed in slightly more detail in Section~\ref{sec:7}. However, the systematic studies presented in refs.~\cite{Csanad:2011jq, Lokos:2024yjm} cannot be considered comprehensive, as they do not examine the logarithmic prefactor discussed, for example, in ref.~\cite{Shen:2013vja}. This prefactor is partly responsible for the fact that the effective temperature of thermal radiation differs from the true temperature even at high temperatures~\cite{Shen:2013vja}. Since the exponential term decays rapidly, the influence of this logarithmic factor is reduced for early-time emission, and therefore, in order to remain within an analytic framework, I neglect it. A thorough investigation of the logarithmic factor certainly deserves further attention, but it lies beyond the scope of the present work; I plan to address it in a future publication.

The present approach emphasizes the space–time distribution of the source, while remaining consistent with the picture of volumetric, instantaneous photon escape from the medium, and it also allows one to remain within an analytical framework throughout the entire calculation.

The invariant momentum distribution of the thermal radiation is derived through a space-time integration of the source function:
\begin{equation}\label{eq:default-integrals}
     \left.\frac{d^2 N}{2\pi p_{\rm T} dp_{\rm T} dy}\right|_{y=0}= \int\limits_{-\infty}^{\infty} d\eta_{z}  \int\limits_{\tau_0}^{\infty} \tau A_{\rm T}(\tau)S\left(x^{\mu},p^{\mu}\right) d\tau  ,
\end{equation}
where $A_{\rm T}$ represents the transverse size, meaning that the above expression has already been integrated over the transverse plane. A simple approximation for the shape of the function $A_{\rm T}(\tau)$ will be discussed later.

Upon substituting the expression of the source function into the integral, several approximations are considered. I make use of the approximation $\Omega \approx \lambda \eta_z$, which holds throughout most of the domain where the solution of Section~\ref{sec:2} is valid. Deviations from linearity appear only near the boundaries of this domain and are negligible compared to its full extent~\cite{Csorgo:2018pxh}. This approximation also corresponds to the $\lambda\rightarrow 1$ limit, and based on fits to experimental data, the value of $\lambda$ is indeed close to unity~\cite{Csorgo:2018pxh,Kasza:2018jtu,Kasza:2018qah}. Ref.~\cite{Csorgo:2018pxh} discusses that as the value of $\lambda$ approaches unity, the domain of validity of the solution presented in Section~\ref{sec:2} broadens accordingly. Although the solution is formally defined on a finite $\eta_z$ interval, the $\lambda\rightarrow 1$ limit causes this interval to expand sufficiently to encompass the physically relevant range of $\eta_z$, especially in the context of the PHENIX detector’s acceptance. Consequently, it does not pose a problem that the integral over $\eta_z$ in eq.~\eqref{eq:default-integrals} extends from $-\infty$ to $+\infty$.

The term $p_{\mu}u^{\mu}$ represents the photon energy in the co-moving frame of the fluid. Near midrapidity ($y\approx 0$), the longitudinal component of the four-momentum can be neglected. Additionally, I make use of the near-unity value of $\lambda$, which implies the approximation $\Omega \approx \lambda \eta_z$:
\begin{equation}
    p_{\mu}u^{\mu} \approx p_{\rm T} \cosh\left(\lambda \eta_z\right).
\end{equation}
Furthermore, by interchanging the order of integration over $\tau$ and $\eta_z$, the hypersurface corresponding to the cessation of thermal photon emission is approximated by a constant average value, $\tau_{\rm f}\left(\eta_z\right)=\tau_{\rm f} = \textnormal{constant}$, in analogy with the treatment of the hypersurface describing the quark-to-hadron transition earlier in the manuscript ($\tau_{\rm tr}\left(\eta_z\right)=\tau_{\rm tr} = \textnormal{constant}$). That approximation will likewise be applied in the forthcoming derivation of thermal radiation.

All this has led to the following integrals, which can also be evaluated analytically:
\begin{equation}\label{eq:spectra_integrals}
    \left.\frac{d^2 N}{2\pi p_{\rm T} dp_{\rm T} dy}\right|_{y=0} \approx \frac{g}{\left(2\pi \hbar\right)^3}\frac{1}{\tau_{\rm R}}\int\limits_{\tau_0}^{\tau_{\rm f}}d\tau\int\limits_{-\infty}^{\infty} d\eta_{z} \: \frac{\tau A_{\rm T}(\tau)\: p_{\rm T} \cosh\left(\lambda \eta_z\right)}{\cosh\left(\left(\lambda-1\right)\eta_z\right)}\exp\left(-\frac{p_{\rm T} \cosh\left(\lambda \eta_z\right)}{T(\tau,\eta_z)}\right),
\end{equation}
where it was employed that the effect of the transparency function $H\left(\tau\right)$ is limited to modifying the upper limit of the integration over $\tau$. Now we reach the stage where the spectrum of thermal radiation is expressed as the sum of contributions from the hadronic and QGP phases. To proceed, we first express the temperature $T$ appearing in the Boltzmann factor as follows:

\begin{align}
    T\left(\tau,\eta_z\right) &= \Theta\left(\tau-\tau_{\rm tr}\right)T_{\rm q}\left(\tau,\eta_z\right) + \Theta\left(\tau_{\rm tr}-\tau\right)T_{\rm h}\left(\tau,\eta_z\right),\\
    T_{\rm q}\left(\tau,\eta_z\right) &= T_{\rm q}\left(\tau,\eta_z;C_{\rm q}, T_{0}, \tau_{0}\right),\\
    T_{\rm h}\left(\tau,\eta_z\right) &= T_{\rm h}\left(\tau,\eta_z;C_{\rm h}, T_{\rm tr,0}, \tau_{\rm tr}\right),
\end{align}
where $T_{\rm q}\left(\tau,\eta_z\right)$ is given by eq.~\eqref{eq:temp-quark}, and $T_{\rm h}\left(\tau,\eta_z\right)$ is defined in eq.~\eqref{eq:temp-hadron}. I employ the matching condition $T_{\rm q}\left(\tau_{\rm tr},\eta_z\right)\approx T_{\rm h}\left(\tau_{\rm tr},\eta_z\right) \approx T_{\rm tr,0}$, which is justified under the approximation that the $\eta_z$-dependence of the hypersurface equation of constant-temperature at $T_{\rm tr,0}$ can be regarded as negligible.
I use a similar approach for the $A_{\rm T}(\tau)$ function, which describes the transverse size. It is expressed in terms of the average transverse size in the QGP phase, $A_{\rm T,q}$, and the average transverse size in the hadronic phase, $A_{\rm T,h}$:
\begin{equation}
    A_{\rm T}(\tau) = \Theta\left(\tau-\tau_{\rm tr}\right)A_{\rm T,q} + \Theta\left(\tau_{\rm tr}-\tau\right)A_{\rm T,h}.
\end{equation}
This formulation of the temperature (and the transverse size) is equivalent to partitioning the integration over $\tau$ at the proper time $\tau_{\rm tr}$, leading to a thermal spectrum composed of distinct high- and low-temperature (i.e., QGP and hadronic) contributions:
\begin{equation}\label{eq:total-spectrum}
    \left.\frac{d^2 N}{2\pi p_{\rm T} dp_{\rm T} dy}\right|_{y=0} \approx \left.\frac{d^2 N_{\rm q}}{2\pi p_{\rm T} dp_{\rm T} dy}\right|_{y=0} + \left.\frac{d^2 N_{\rm h}}{2\pi p_{\rm T} dp_{\rm T} dy}\right|_{y=0}.
\end{equation}
The component originating from the QGP is labeled with the subscript "q", while the contribution from the hadronic phase is denoted by the subscript "h". The explicit expressions of the two components are given as follows:

\begin{align}
\left.\frac{d^2 N_{\rm q}}{2\pi p_{\rm T} dp_{\rm T} dy}\right|_{y=0} &= \frac{g}{\left(2\pi \hbar\right)^3}\frac{1}{\tau_{\rm R}}\int\limits_{\tau_0}^{\tau_{\rm tr}}d\tau\int\limits_{-\infty}^{\infty} d\eta_{z} \: \frac{\tau A_{\rm T,q}\: p^{\mu}u_{\mu}}{\cosh\left(\Omega-\eta_z\right)}\exp\left(-\frac{p^{\mu}u_{\mu}}{T_{\rm q}\left(\tau,\eta_z\right)}\right),\label{eq:integral-final-quark}\\
\left.\frac{d^2 N_{\rm h}}{2\pi p_{\rm T} dp_{\rm T} dy}\right|_{y=0} &= \frac{g}{\left(2\pi \hbar\right)^3}\frac{1}{\tau_{\rm R}}\int\limits_{\tau_{\rm tr}}^{\tau_{\rm f}}d\tau\int\limits_{-\infty}^{\infty} d\eta_{z} \: \frac{\tau A_{\rm T,h}\: p^{\mu}u_{\mu}}{\cosh\left(\Omega-\eta_z\right)}\exp\left(-\frac{p^{\mu}u_{\mu}}{T_{\rm h}\left(\tau,\eta_z\right)}\right).\label{eq:integral-final-hadron}
\end{align}
The analytic evaluation of the integrals was carried out using the saddle-point approximation. The detailed steps of the derivation are discussed in the Appendix of ref.~\cite{Kasza:2023rpx}, where I performed the same calculations for the thermal photon spectrum as in the present model, the only difference being that a single-component model was considered there. After evaluating the integrals in eqs.~\eqref{eq:integral-final-quark} and~\eqref{eq:integral-final-hadron}, the following results were obtained for the components of QGP and hadronic matter:
\begin{align}
    \left.\frac{d^2 N_{\rm q}}{2\pi p_{\rm T} dp_{\rm T} dy}\right|_{y=0}&=N_{\rm 0,q}\:\frac{2\alpha_{\rm q}}{3\pi^{3/2}}\left[\frac{1}{T_{\rm tr}^{\alpha_{\rm q}}}-\frac{1}{T_{0}^{\alpha_{\rm q}}}\right]^{-1} p_{\rm T}^{-\alpha_{\rm q}-2}\left.\Gamma\left(\alpha_{\rm q}+\frac{5}{2},\frac{p_{\rm T}}{T}\right)\right|^{T=T_0}_{T=T_{\rm tr}},\label{eq:quark-component}
\\
    \left.\frac{d^2 N_{\rm h}}{2\pi p_{\rm T} dp_{\rm T} dy}\right|_{y=0}&=N_{\rm 0,h}\:\frac{2\alpha_{\rm h}}{3\pi^{3/2}}\left[\frac{1}{T_{\rm f}^{\alpha_{\rm h}}}-\frac{1}{T_{\rm tr}^{\alpha_{\rm h}}}\right]^{-1} p_{\rm T}^{-\alpha_{\rm h}-2}\left.\Gamma\left(\alpha_{\rm h}+\frac{5}{2},\frac{p_{\rm T}}{T}\right)\right|^{T=T_{\rm tr}}_{T=T_{\rm f}},\label{eq:hadronic-component}
\end{align}
where the midrapidity density $N_{0,i}$ for $i \in \{{\rm q},{\rm h}\}$ is calculated as follows:
\begin{equation}\label{eq:N0_integral}
    N_{0,i}=\left.\frac{dN_i}{dy}\right|_{y=0}=\int\limits_{0}^{\infty} \left.\frac{d^2 N_i}{dp_{\rm T} dy}\right|_{y=0} dp_{\rm T}, 
\end{equation}
and performing the integral for $\forall i$ yielded the following results:
\begin{align}
    N_{\rm 0,q}&=\frac{g A_{\rm T,q}}{\left(2\pi \hbar\right)^3}\frac{\tau_0}{\tau_{\rm R}}\frac{T_0^{\alpha_{\rm q}+3}}{\alpha_{\rm q}} \left[\frac{1}{T_{\rm tr}^{\alpha_{\rm q}}}-\frac{1}{T_{0}^{\alpha_{\rm q}}}\right]\frac{3\pi^{3/2}C_{\rm q}}{2\lambda}\left(\frac{2\pi C_{\rm q}}{\lambda^2\left(2C_{\rm q}-1\right)-\lambda\left(C_{\rm q}-1\right)}\right)^{1/2},\\
    N_{\rm 0,h}&=\frac{g A_{\rm T,h}}{\left(2\pi \hbar\right)^3}\frac{\tau_{\rm tr}}{\tau_{\rm R}}\frac{T_{\rm tr}^{\alpha_{\rm h}+3}}{\alpha_{\rm h}} \left[\frac{1}{T_{\rm f}^{\alpha_{\rm h}}}-\frac{1}{T_{\rm tr}^{\alpha_{\rm h}}}\right]\frac{3\pi^{3/2}C_{\rm h}}{2\lambda}\left(\frac{2\pi C_{\rm h}}{\lambda^2\left(2C_{\rm h}-1\right)-\lambda\left(C_{\rm h}-1\right)}\right)^{1/2}.   
\end{align}
In the expression describing the direct photon spectrum, several new notations have been introduced, which correspond to the following definitions:
\begin{align}
    T_{\rm f} &= T_{\rm tr,0} \left(\frac{\tau_{\rm tr}}{\tau_{\rm f}} \right)^{\frac{\lambda}{C_{\rm h}}},\\
    T_{\rm tr} &= T_{0} \left(\frac{\tau_{0}}{\tau_{\rm tr}} \right)^{\frac{\lambda}{C_{\rm q}}},    
\end{align}
thus, as a consequence of the manner in which the two solutions are matched, we have  $T_{\rm tr} = T_{\rm tr,0}$. The parameters $\alpha_{\rm h}$ and $\alpha_{\rm q}$ were also introduced, defined by the following relations for $i \in \{{\rm q},{\rm h}\}$:
\begin{equation}\label{eq:alpha_i}
    \alpha_i = \frac{2C_i}{\lambda} - 3.
\end{equation}
The structure of eqs.~\eqref{eq:quark-component} and~\eqref{eq:hadronic-component} reveals that the equation-of-state parameters $C_{\rm q}$ and $C_{\rm h}$ effectively scale out of the thermal photon spectrum. As a result, their values cannot be uniquely determined through fitting to experimental data unless some other model parameters are fixed.\footnote{To illustrate this concept with a simple example: if one fits the function $f(x) = \frac{A}{B}x$ to data, the fit will only constrain the ratio $A/B$. In such a case, the individual value of parameter $A$ can only be determined if we fix $B$ to some value.} In contrast, temperature-related quantities, such as the initial temperature $T_0$, the average transition temperature $T_{\rm tr}$, and the average temperature $T_{\rm f}$ (not identical to $T_{\rm fo}$) that characterize the cessation of thermal photon production, can be constrained by comparison with experimental measurements. This will be the focus of the following section.

\section{Comparison to PHENIX data}\label{sec:4}
The PHENIX Collaboration has recently measured the non-prompt component of direct photons in $Au+Au$ collisions at a center-of-mass energy of $\sqrt{s_{NN}} = 200$ GeV. This was achieved by scaling the fit to the $p+p$ data by the number of binary collisions, which provides an estimate of the prompt contribution arising from hard scattering processes~\cite{PHENIX:2022rsx}. This prompt yield was then subtracted from the direct photon spectrum. The resulting non-prompt component is understood to be predominantly of thermal origin and can therefore be effectively described using hydrodynamic models, such as the one presented in Sections 2 and 3. At this point, I emphasize that, to remain within a fully analytic framework, my model incorporates several approximations in the determination of the thermal emission spectrum. Consequently, the model is suitable only for qualitative analysis, even though in the comparison with data I will employ tools characteristic of quantitative studies, so as to extract as much information as possible for the later refinement and improvement of the model. Nonetheless, the reader should be cautioned that any predictions extracted from this model for the physical parameters should not be interpreted as quantitatively robust.

The PHENIX Collaboration measured the non-prompt photon contribution in multiple centrality classes, specifically in the 0–20\%, 20–40\%, 40–60\%, and 60–93\% intervals. \linebreak I compare my model to the data in all four centrality classes; however, it is important to emphasize that the validity of the model becomes increasingly questionable in the two most peripheral classes. This limitation can be traced back to several underlying reasons.

Hydrodynamics inherently assumes that the medium is at least locally in thermal equilibrium. In peripheral collisions, the number of participating nucleons is smaller, resulting in a medium with lower energy and particle density. In addition, the geometry becomes more asymmetric, leading to stronger gradients and rendering the establishment of local thermal equilibrium more uncertain~\cite{STAR:2004jwm, PHENIX:2006dpn, ALICE:2010suc, Voloshin:1994mz,STAR:2000ekf,Sahu:2002sp,PHENIX:2009cjr, Richardson:2012kq,Yadav:2025vtc}. This expectation is supported by hydrodynamic simulations (e.g., MUSIC~\cite{Ryu:2012at}, VISH2+1~\cite{Song:2007fn,Song:2007ux}), which qualitatively describe the general features of photon spectra in central collisions~\cite{Shen:2013vja,Paquet:2015lta}, although quantitative discrepancies persist to some extent. In peripheral events, both the thermal photon yield and the flow coefficients (such as $v_2$) are generally not described as accurately, indicating a breakdown of equilibrium assumptions. Furthermore, in such small systems, pre-equilibrium dynamics (e.g., Glasma or far-from-equilibrium QCD effects) can dominate over a longer time scale. This implies a relatively larger contribution from early-stage, non-thermal photons, thereby diminishing the role of hydrodynamics. Importantly, in the context of my model — which is based on perfect fluid description — viscous corrections are expected to play a more significant role in peripheral collisions. This is due to the stronger velocity field gradients that naturally arise in smaller systems.

During the fitting procedure, certain model parameters were fixed, while others were constrained within physically motivated intervals guided by constraints derived from lattice QCD simulations. This approach was adopted in order to assess whether the hydrodynamic model is capable of describing the thermal photon spectrum using parameter values that are consistent with lattice QCD results. The vast majority of lattice QCD simulations are performed at zero or very small baryon chemical potential, which ensures the precision and reliability of the calculations. This does not pose a problem in our case, as results from the STAR collaboration have shown that, at chemical freeze-out, there is almost an order-of-magnitude difference between the temperature and the baryon chemical potential~\cite{STAR:2017sal}.

The value of $T_{\rm tr}$ was fixed at 156 MeV, in accordance with simulations performed by the Budapest–Wuppertal collaboration, which identify this temperature as the characteristic value associated with the QCD cross-over transition from hadronic matter to quark–gluon plasma~\cite{Borsanyi:2010cj}. Importantly, this cross-over temperature does not represent a sharp phase boundary but rather a temperature region in which hadronic matter begins to dissolve into its partonic constituents. It is often associated with the so-called Hagedorn temperature; however, the Hagedorn temperature is not a true phase transition temperature in the conventional thermodynamic sense, but rather a statistical property of the hadronic spectrum. Consequently, its reported values span a broad range in the literature~\cite{Hagedorn:1965st, Broniowski:2000bj,Broniowski:2004yh,Cohen:2011cr,Cleymans:2011fx}.

The value of $\alpha_{\rm q}$ was also fixed, noting that the parameter $C_{\rm q}$ is equal to the high-temperature limit of $\kappa_{\rm q}$, which corresponds to the conformal limit and yields $C_{\rm q} = 3$. Moreover, the parameter $\lambda$ can be set between 1.1 and 1.3, based on studies of hadronic observables, for which these values are considered optimal~\cite{Kasza:2018qah,Kasza:2018jtu}. Under these considerations, I have fixed the value of $\alpha_{\rm q}$ at 2, in accordance with eq.~\eqref{eq:alpha_i}.

While fitting the model to the data, the value of $\alpha_{\rm h}$ was constrained within an interval, the bounds of which were determined based on the parameter $C_{\rm h}$. This parameter corresponds to the high-temperature limit of the function $\kappa_{\rm h}$ and also controls the peak value of the temperature-dependent function $\kappa_{\rm tot}(T)$ defined in eq.~\eqref{eq:total-eos}. The temperature dependence of $\kappa$ predicted by lattice QCD can be readily computed using the tables provided in ref.~\cite{Borsanyi:2010cj}. This calculation was carried out together with a co-author in ref.~\cite{Csorgo:2016ypf}, where it is clearly shown that the maximum of $\kappa$ — which occurs around $T \approx 156$ MeV — is approximately 7. However, in this temperature range ($T < 170$ MeV), the uncertainties of the lattice QCD results increase significantly. For this reason, I assigned a wider range for $C_{\rm h}$, and correspondingly for $\alpha_{\rm h}$, specifically choosing $\alpha_{\rm h} \in \{6,11\}$. For $\lambda \approx 1.3$, this corresponds to the interval $C_{\rm h} \in \{5.9,9.1\}$ based on eq.~\eqref{eq:alpha_i}, implying that the peak of the $\kappa_{\rm tot}(T)$ function can vary approximately between 5.5 and \nolinebreak[4]8, which also depends on the chosen values of $T_{\rm fo}$ and $\kappa_{\rm fo}$.

During the fitting procedure, the parameters $T_0$, $N_{\rm 0,h}$, and $N_{\rm 0,q}$ were treated as free parameters; however, the available data do not allow the model to make a meaningful prediction for the value of $N_{\rm 0,h}$. In the initial trials, only an upper limit was imposed on the parameter $T_{\rm f}$, since it logically cannot exceed the transition temperature $T_{\rm tr}$. It was further observed that $T_{\rm f}$ tends to converge towards $T_{\rm tr}$ during the fits, suggesting that the hadronic phase contribution becomes negligible. Correspondingly, in these cases $N_{\rm 0,h}$ approaches zero. However, as subsequent fitting results indicate, the experimental data lack sufficient information to accurately determine $N_{\rm 0,h}$, leading to numerical instabilities in the fitting process. Due to this, each dataset was fitted using two different methods. In one case, the value of $N_{\rm 0,h}$ was fixed to zero (which is equivalent to fixing the value of $T_{\rm f}$ at $T_{\rm tr}=156$ MeV), thus the entire dataset was described using the component of quark matter (labeled as QGP component in the figures) without considering the hadronic phase. In the other case, the value of $T_{\rm f}$ was fixed to 150 MeV, close to $T_{\rm tr}$. This ensured that the hadronic contribution originates only from a narrow temperature range, which however provides a non-negligible contribution to the low-$p_{\rm T}$ part of the thermal photon yield.

\begin{figure*}[!h]
\subfloat[Hadronic Source Included]{\includegraphics[width=0.49\linewidth]{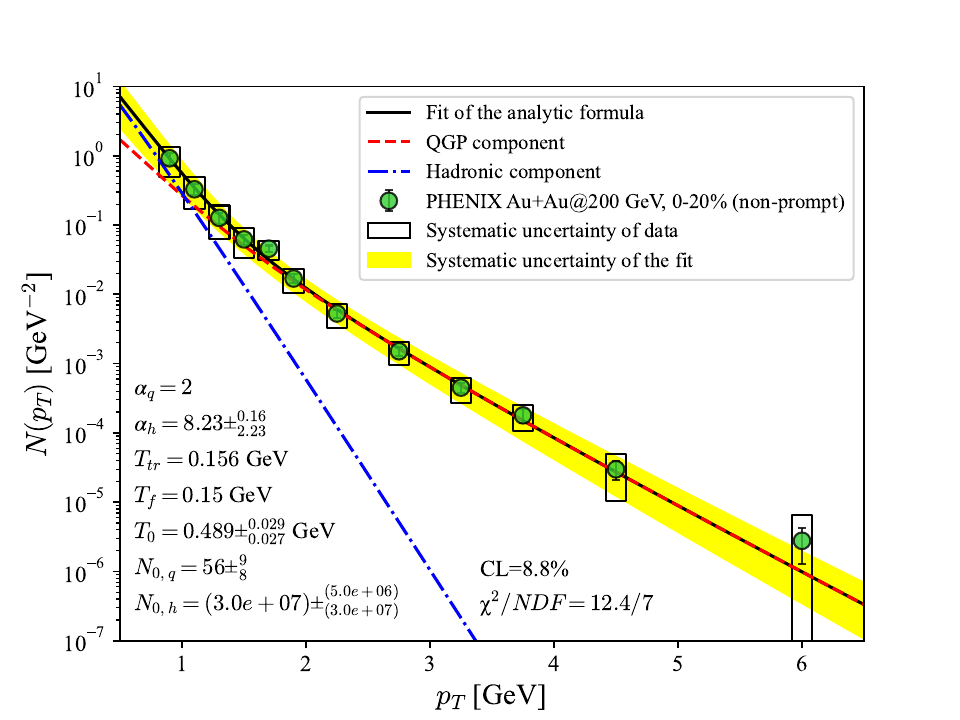}}
\hfil
\subfloat[Hadronic Source Excluded]{\includegraphics[width=0.49\linewidth]{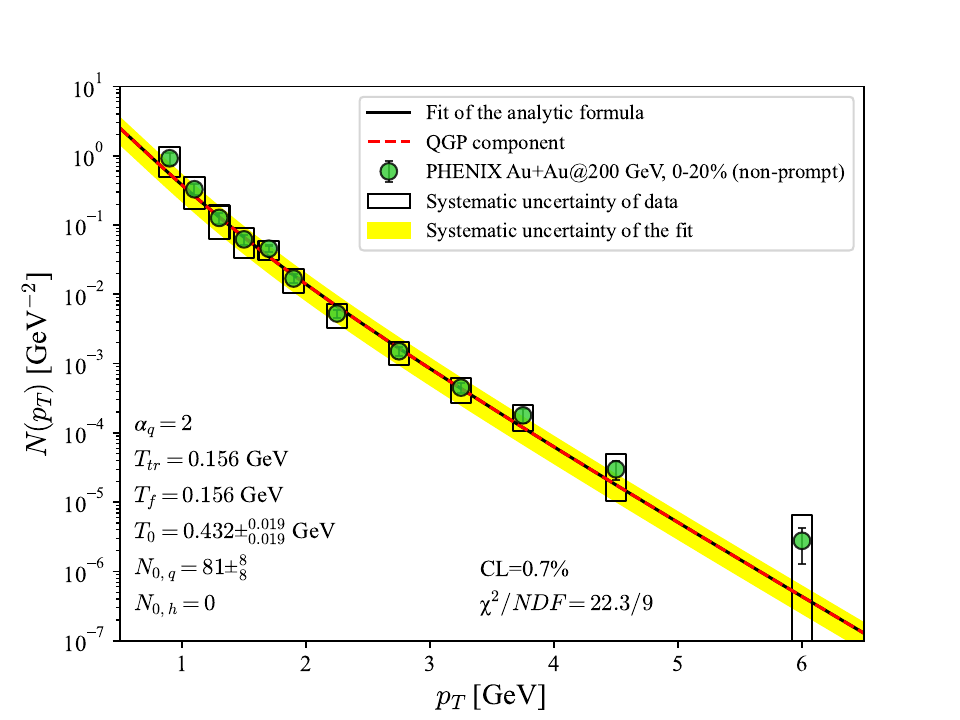}}
\caption{The fit of equation~\eqref{eq:total-spectrum} to the non-prompt direct photon spectrum measured by the PHENIX collaboration in $Au+Au$ collisions at $\sqrt{s_{\rm NN}} =200$ GeV for the 0–20\% centrality class. The measured data points are indicated by green markers. The blue dot-dashed curve depicts the hadronic component, as defined by eq.~\eqref{eq:hadronic-component}. The red dashed curve corresponds to the yield of QGP, whose analytical expression is provided by eq.~\eqref{eq:quark-component}. The sum of these two contributions yields the total spectrum, represented by the solid black curve. The error bars denote the statistical errors of the experimental data, whereas the boxes indicate the systematic uncertainties of the same. The yellow band represents the systematic uncertainty of the model fit, quantified based on the systematic uncertainties propagated from the fitted model parameters.}
\label{fig:fig1}
\end{figure*}

\begin{figure*}[!h]
\subfloat[Hadronic Source Included]{\includegraphics[width=0.49\linewidth]{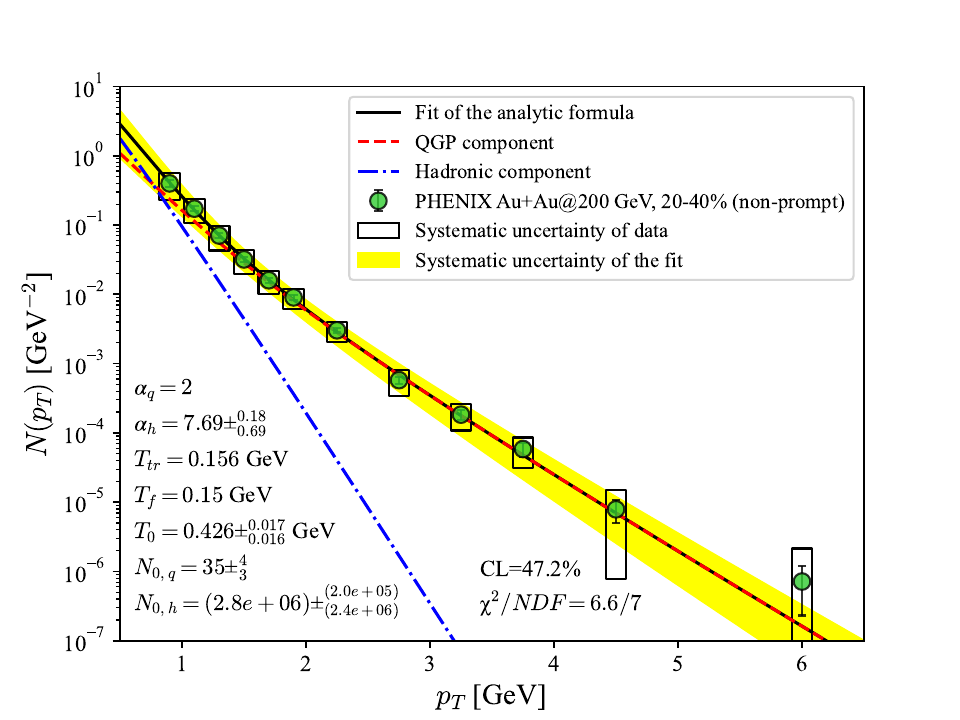}%
\label{fig_first_case}}
\hfil
\subfloat[Hadronic Source Excluded]{\includegraphics[width=0.49\linewidth]{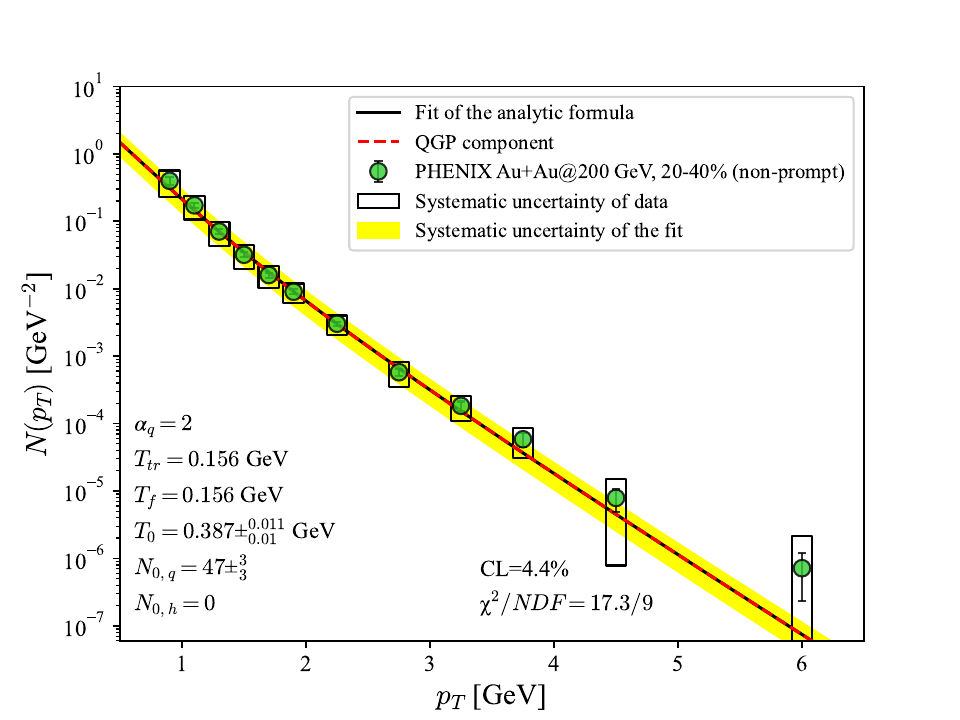}%
\label{fig_second_case}}
\caption{The fit of equation~\eqref{eq:total-spectrum} to the non-prompt direct photon spectrum measured by the PHENIX collaboration in $Au+Au$ collisions at $\sqrt{s_{\rm NN}} =200$ GeV for the 20–40\% centrality class. The curves, error bars, and uncertainty boxes presented in this figure follow the same notation scheme as those used in Fig.~\ref{fig:fig1}.}
\label{fig:fig2}
\end{figure*}

\begin{figure*}[!h]
\subfloat[Hadronic Source Included]{\includegraphics[width=0.49\linewidth]{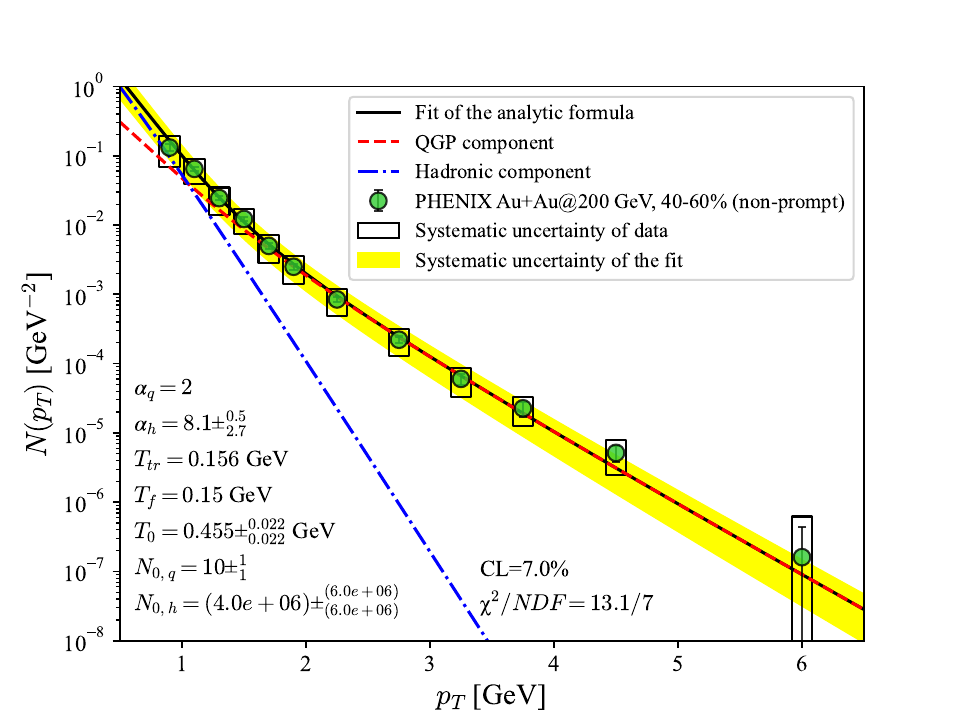}%
\label{fig_first_case}}
\hfil
\subfloat[Hadronic Source Excluded]{\includegraphics[width=0.49\linewidth]{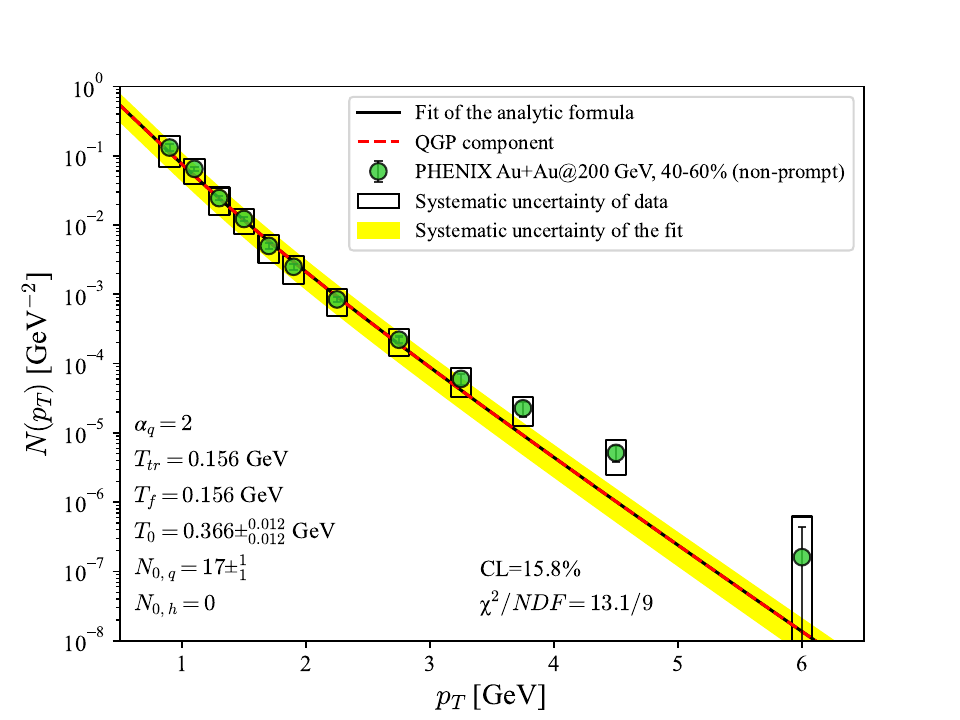}%
\label{fig_second_case}}
\caption{The fit of equation~\eqref{eq:total-spectrum} to the non-prompt direct photon spectrum measured by the PHENIX collaboration in $Au+Au$ collisions at $\sqrt{s_{\rm NN}}=200$ GeV for the 40–60\% centrality class.The curves, error bars, and uncertainty boxes presented in this figure follow the same notation scheme as those used in Fig.~\ref{fig:fig1}.}
\label{fig:fig3}
\end{figure*}

\begin{figure*}[!h]
\subfloat[Hadronic Source Included]{\includegraphics[width=0.49\linewidth]{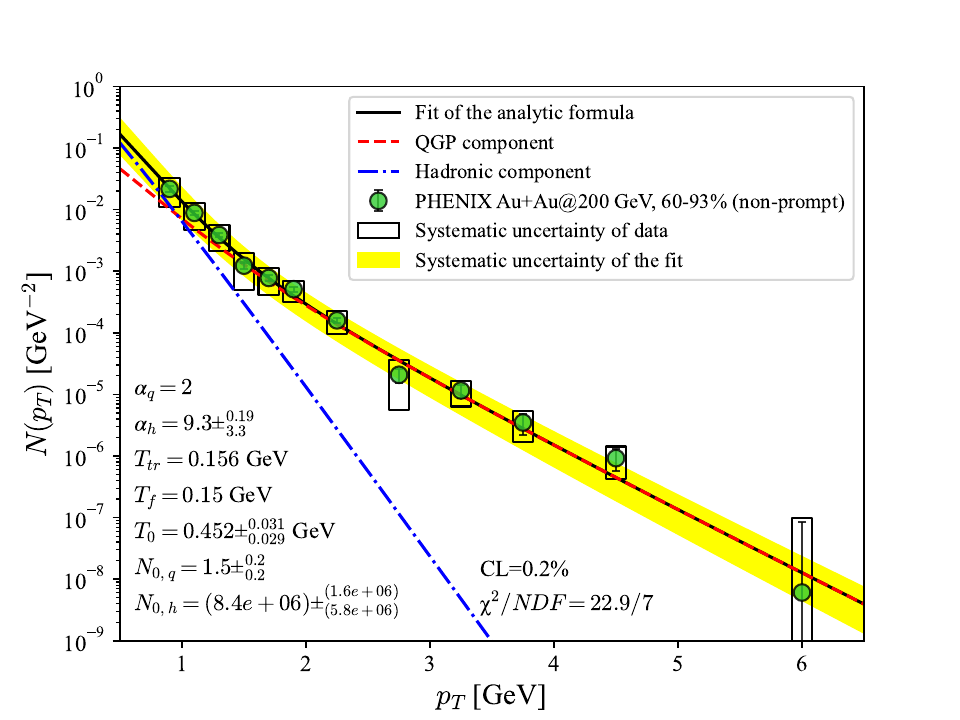}%
\label{fig_first_case}}
\hfil
\subfloat[Hadronic Source Excluded]{\includegraphics[width=0.49\linewidth]{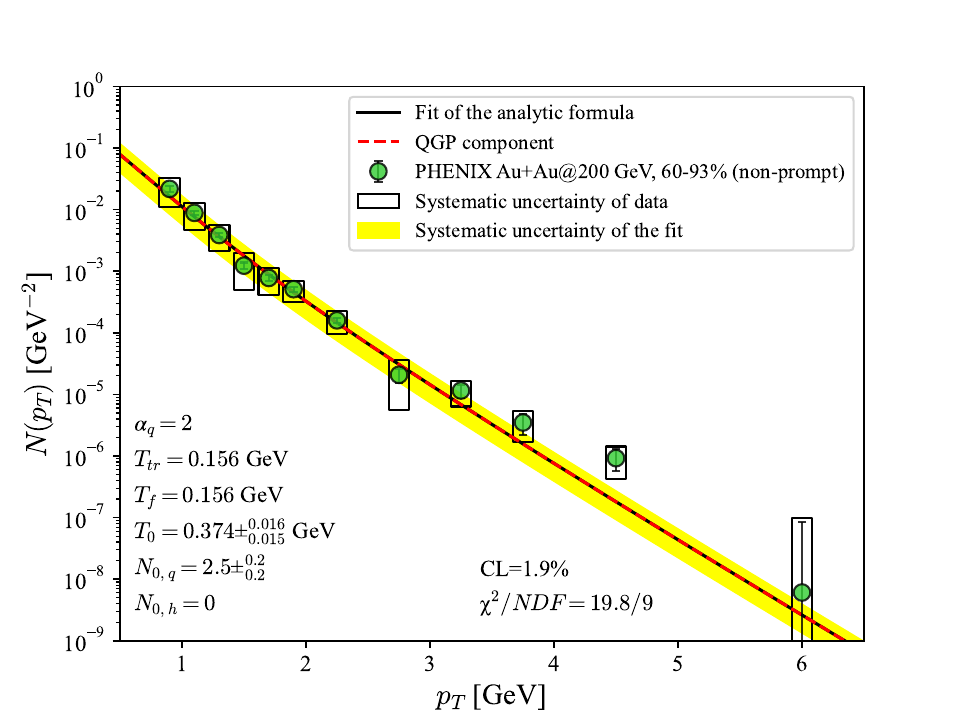}%
\label{fig_second_case}}
\caption{The fit of equation~\eqref{eq:total-spectrum} to the non-prompt direct photon spectrum measured by the PHENIX collaboration in $Au+Au$ collisions at $\sqrt{s_{\rm NN}}=200$ GeV for the 60–93\% centrality class. The curves, error bars, and uncertainty boxes presented in this figure follow the same notation scheme as those used in Fig.~\ref{fig:fig1}.}
\label{fig:fig4}
\end{figure*}

The comparison between the experimental data and the model is presented in Figures~\ref{fig:fig1}-\ref{fig:fig4}. In Figure~\ref{fig:fig1}, the model introduced in Section~\ref{sec:3} is applied to the data of the centrality class 0–20\%. Figures~\ref{fig:fig2} through~\ref{fig:fig4} extend this comparison to the 20–40\%, 40–60\%, and 60–93\% centrality intervals, respectively. In each figure, the left panel shows the fit including both the hadronic and QGP components, whereas the right panel displays the fit considering only the thermal photons originating from the QGP (thus, the black solid and red dashed lines coincide in the right panels). The fitted parameter values and their statistical errors are also presented in the figures.

The confidence level (CL) associated with each fit is indicated in the corresponding figures. In all cases, I find $\textnormal{CL}>0.1\%$ implying that the deviation between the model and the data remains within the 3$\sigma$ range. The least favorable fit yields a confidence level of 0.2\%, while the best-fit scenario reaches 47.2\%. The CL values suggest that the two-component model provides a better description of the two most central data sets, whereas in the more peripheral classes, improved agreement with the data is achieved when the hadronic contribution is omitted.

When thermal radiation from the hadronic phase is excluded, the extracted initial temperature $T_0$ shows a decreasing trend with centrality, except for the 60–93\% class, where the central value of $T_0$ is higher than in the 40–60\% class. However, considering the statistical errors, neither a plateau nor a mild decrease in $T_0$ can be excluded.

In contrast, when both the QGP and the hadronic components are included in the thermal photon yield, no such monotonic trend is observed: the minimum value of the initial temperature occurs in the 20–40\% centrality class. Even after accounting for statistical uncertainties, the data are consistent with a roughly constant initial temperature at centralities above 20\%.

\begin{figure}[h!]
    \centering
    \includegraphics[width=0.74\linewidth]{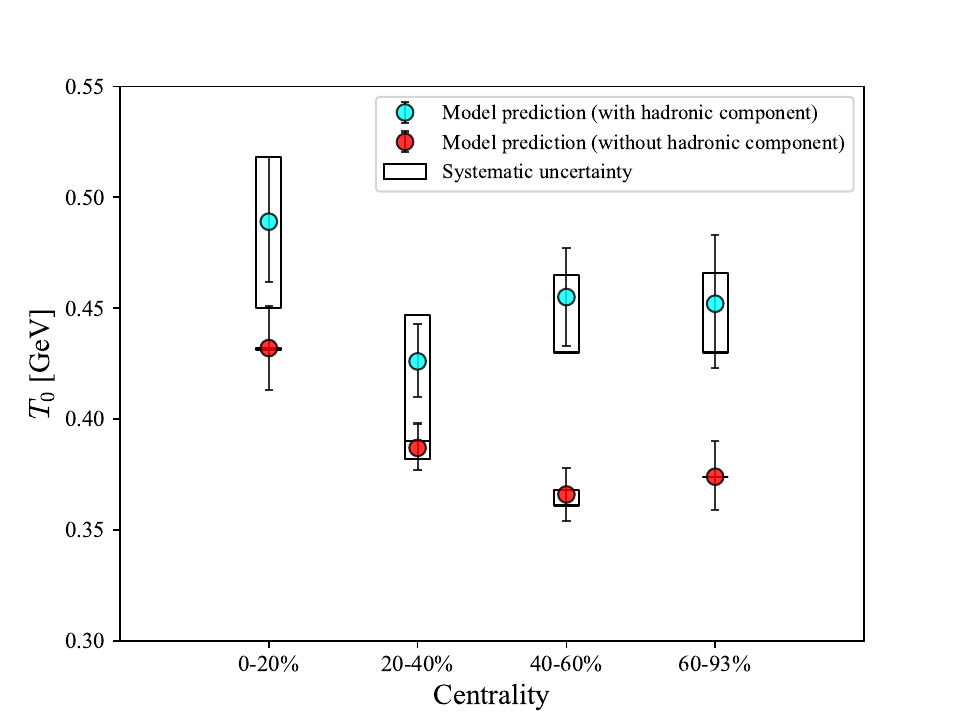}
    \caption{Extracted initial temperatures as a function of centrality, obtained from comparisons between the experimental data and my analytic model with (cyan) and without (red) the inclusion of the hadronic component, shown here as qualitatively indicative results.}
    \label{fig:fig5}
\end{figure}

To aid transparency, Table~\ref{tab:tab1} summarizes and Figure~\ref{fig:fig5} illustrates the extracted values of the initial temperature $T_0$ for each centrality class, along with the corresponding systematic uncertainties. However, because of the reasons discussed at the beginning of this section, these results should be regarded as qualitatively indicative and should not be considered quantitatively precise. It should also be noted that the results for the two most peripheral centrality classes warrant particular caution, given the limited validity of the hydrodynamic description in these regimes, as discussed previously.

\begin{table}[!t]
    \centering
    \renewcommand{\arraystretch}{1.2}
    \begin{tabular}{c|c|c}
        \hline \multicolumn{3}{c}{Hadronic source included}\\ \hline
         Centrality [\%] & $T_0\pm(\textnormal{stat.})\pm(\textnormal{syst.})$ [MeV] & CL [\%]\\ \hline
         0 - 20 & $489 \pm^{29}_{27}\pm^{29}_{39}$ & 8.8\\
         20 - 40 & $426 \pm^{17}_{16}\pm^{21}_{44}$ & 47.2\\
         40 - 60 & $455 \pm 22\pm^{10}_{25}$ & 7.0 \\
         60 - 93 & $452\pm^{31}_{29}\pm^{14}_{22}$ & 0.2 \\ \hline
         \multicolumn{3}{c}{Hadronic source excluded}\\ \hline
         Centrality [\%] & $T_0\pm(\textnormal{stat.})\pm(\textnormal{syst.})$ [MeV] & CL [\%]\\ \hline
         0 - 20 & $432\pm 19 \pm^{0}_{1}$& 0.7\\
         20 - 40 & $387 \pm^{11}_{10}\pm^{3}_{5}$ & 4.4\\
         40 - 60 & $366 \pm 12\pm^{2}_{5}$ & 15.8 \\
         60 - 93 & $374 \pm^{16}_{15}\pm 0$ & 1.9 \\ \hline         
    \end{tabular}
    \caption{Extracted initial temperatures and corresponding confidence levels (CL) as a function of centrality, obtained from comparisons between the experimental data and my analytic model with (upper rows) and without (lower rows) the inclusion of the hadronic component, summarized here as qualitatively indicative results.}
    \label{tab:tab1}
\end{table}

\clearpage

\clearpage

\section{Validation of the Model in the Hadronic Channel}\label{sec:5}

In Section~\ref{sec:4}, I demonstrated that the model describes the overall behavior of the non-prompt direct photon spectra observed by PHENIX. However, it is not yet clear whether the obtained parameter set can reproduce the trends observed in hadronic measurements.

The 1+1 dimensional hydrodynamic model applied describes purely longitudinal expansion and therefore does not include transverse flow. Since the hadronic $p_{\rm T}$ spectra are strongly affected by the development of transverse dynamics, the model is not capable of providing a reliable quantitative description in this respect: in such a framework the $p_{\rm T}$ spectra could only be defined through embedding the model into 1+3 dimensions, thus the effect of radial flow would not appear. For this reason only the pseudorapidity distributions of hadrons can be fitted, which primarily reflect the imprint of the longitudinal dynamics and therefore constitute a natural quantity for comparison within the framework of the 1+1 dimensional model.

To test whether the model, with the parameter set obtained in Section~\ref{sec:4}, can provide a satisfactory description of the hadronic channel, I will compare the model with the pseudorapidity densities $dN/d\eta$ measured by the PHOBOS Collaboration in 200 GeV $Au+Au$ collisions~\cite{PHOBOS:2010eyu}. Since PHOBOS reported these data only in central and mid-central collisions, the comparison will be restricted to the 0–20\% and 20–40\% centrality classes. Fortunately, these centrality classes are particularly relevant for testing, since, as noted earlier, the hydrodynamic picture becomes less reliable in more peripheral collisions.

In refs.~\cite{Csorgo:2018pxh,Kasza:2018qah,Csorgo:2018fbz}, my collaborators and I presented a parametric formula for describing the pseudorapidity densities, derived from the solution published in ref.~\cite{Csorgo:2018pxh}. This is the same solution whose generalization was introduced in Section \ref{sec:solution}. In Refs.~\cite{Csorgo:2018pxh,Kasza:2018qah}, the hadronic observables were calculated using the same solution, methods, and similar approximations as those employed here for thermal radiation, with the only distinction that the hadronic channel was considered. This demonstrates that the present model framework is suitable for the investigation of both hadronic and photonic channels. The formula reported in refs.~\cite{Csorgo:2018pxh,Kasza:2018qah}, however, requires a modification in the sense that it should now be derived from the extended solution discussed in Section~\ref{sec:solution}. As hadronic observables reflect the conditions at kinetic freeze-out, the corresponding calculation is, on the one hand, considerably simpler than in the case of thermal photons, and on the other hand, the resulting formula involves only parameters characterizing hadrons. The detailed derivation lies beyond the scope of the present work\footnote{In the course of the derivation, I employed the same methods and followed the same steps as those described in more detail in refs.~\cite{Csorgo:2018pxh,Kasza:2018jtu}.}, and therefore only the final result is presented here, as a parametric curve:
\begin{equation}
	\left(\eta(y),\:\frac{dN}{d\eta}(y)  \right) =
	\left(\eta(y),\: \mathcal{J}(y)\frac{dN}{dy} \right).
	\label{eq:dndeta}
\end{equation}
Here, the Jacobian factor $\mathcal{J}(y)$ and the pseudorapidity $\eta$ can be written as follows~\cite{Kasza:2018jtu}:
\begin{align}
	\mathcal{J}(y) &= \sqrt{1-\frac{m^2}{\bar{m}_{\rm T}(y)^2 \cosh^2(y)}},\label{eq:J-y} \\
    \eta(y)  &=  \tanh^{-1} \left[ \frac{\tanh(y)}{\mathcal{J}(y)} \right],\label{eq:eta-y}    
\end{align}
and both are expressed in terms of the rapidity-dependent average transverse mass~\cite{Kasza:2018qah}:
\begin{equation}
	\bar{m}_{\rm T}(y) \approx m + \frac{T_{\rm eff}}{\cosh^{\alpha\left(C_{\rm h},\lambda\right)}\left(\frac{y}{\alpha(1,\lambda)}\right)}, \label{eq:mtbar-y}
	\end{equation}
where $m$ is the particle mass and $T_{\rm eff}$ stands for the effective temperature, i.e., the inverse slope parameter of the hadronic $p_{\rm T}$ spectrum. The function $\alpha\left(C_{\rm h},\lambda\right)$ is distinct from both $\alpha_{\rm h}$ and $\alpha_{\rm q}$, and is defined as: $\alpha\left(C_{\rm h},\lambda\right) = \left(2\lambda - C_{\rm h}\right)/\left(\lambda-C_{\rm h}\right)$. Finally, the rapidity density in eq.~\eqref{eq:dndeta} is given by:
\begin{equation}
	\frac{dN}{dy} \approx
	\left.\frac{dN}{dy}\right|_{y=0} 
	\cosh^{-\frac{1}{2}\alpha\left(C_{\rm h},\lambda\right)-1}\left(\frac{y}{\alpha(1,\lambda)}\right)
	\exp\left(-\frac{m}{T_{\rm eff}} 
	\left[\cosh^{\alpha\left(C_{\rm h},\lambda\right)}\left(\frac{y}{\alpha(1,\lambda)}\right)-1\right]\right).\label{eq:dndy-function}
\end{equation}
The fitting of eq.~\eqref{eq:dndeta} to the data involves the determination of five parameters. The particle mass is fixed at $m=140$ MeV under the assumption of pion dominance. The rapidity density at midrapidity, $\left.dN/dy\right|_{y=0}$, and the effective temperature, $T_{\rm eff}$, are treated as free fitting parameters. The acceleration parameter of the local velocity field, $\lambda$, together with the hadronic equation-of-state parameter, $C_{\rm h}$, are constrained by the fit to the non-prompt direct photon spectra. The parameters $\lambda$ and $C_{\rm h}$ must take values such that, according to eq.~\eqref{eq:alpha_i}, they reproduce the values of $\alpha_{\rm h}$ shown in Figs.~\ref{fig:fig1} and~\ref{fig:fig2}. The question, therefore, is whether the PHOBOS pseudorapidity density data can be described with values of $\lambda$ and $C_{\rm h}$ that are consistent with the $\alpha_{\rm h}$ values obtained from the fit to the non-prompt direct photon spectra, while at the same time yielding physically acceptable values for the parameters $T_{\rm eff}$ and $\left.dN/dy\right|_{y=0}$. Obviously, to address this question, only those fits from Section~\ref{sec:4} are relevant in which the hadronic source of thermal photons was not neglected.

\begin{figure*}[!h]
\subfloat[0-20\%]{\includegraphics[width=0.49\linewidth]{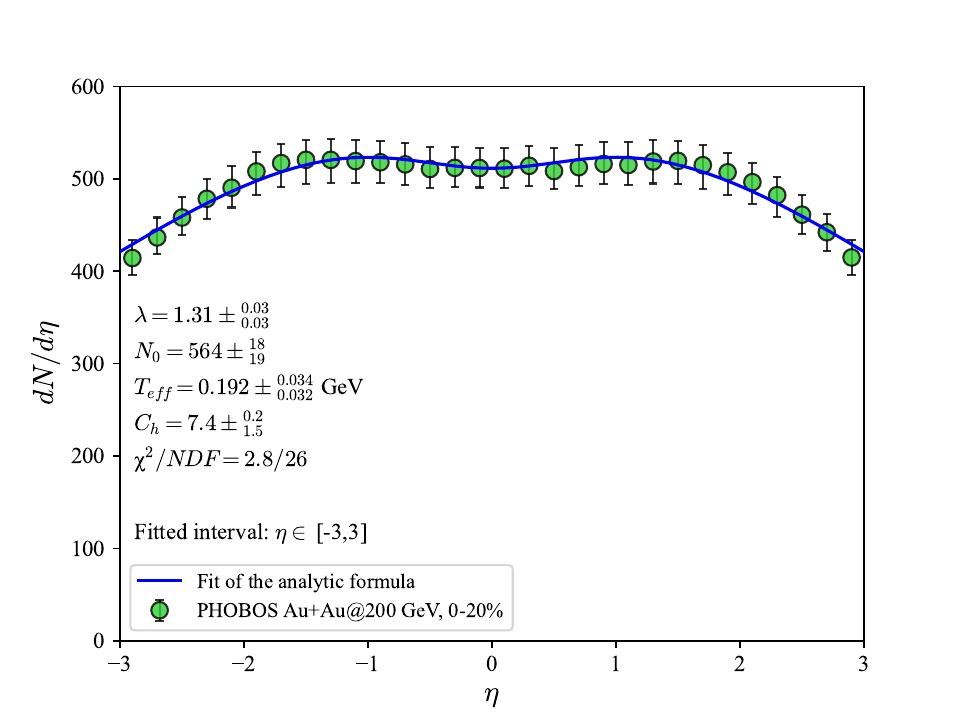}}
\hfil
\subfloat[20-40\%]{\includegraphics[width=0.49\linewidth]{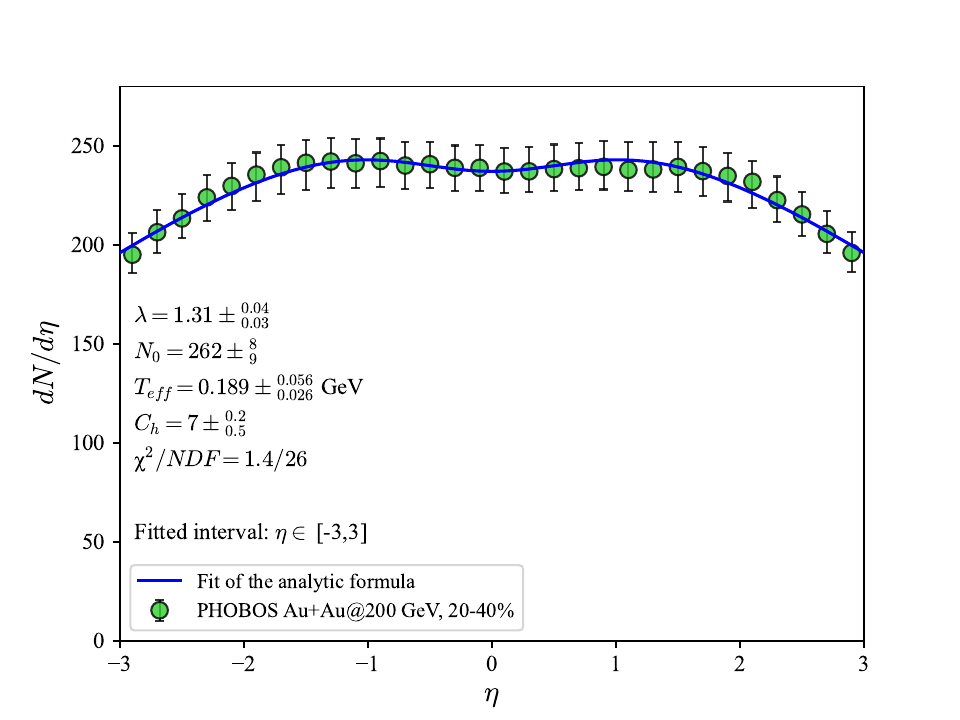}}
\caption{
The fit of equation~\eqref{eq:dndeta} to the pseudorapidity density measured by the PHOBOS collaboration in $Au+Au$ collisions at $\sqrt{s_{\rm NN}} =200$ GeV for the 0–20\% and 20-40\% centrality classes~\cite{PHOBOS:2010eyu}. The measured data points are indicated by green markers. The blue solid line corresponds to the model prediction, whose analytical expression is provided by eqs.~\eqref{eq:dndeta}-\eqref{eq:dndy-function}. The error bars denote the published errors of the experimental data. The parameter $N_0$ represents $\left.dN/dy\right|_{y=0}$.
}
\label{fig:fig6}
\end{figure*}

The model was fitted to the PHOBOS data from 200 GeV $Au+Au$ collisions in the 0–20\% and 20–40\% centrality classes, as illustrated in Fig.~\ref{fig:fig6}. The fits were carried out within a narrower pseudorapidity range, $\eta \in [-3,3]$, than the full extent of the data set, since the solution presented in Section~\ref{sec:solution} is defined only on a finite domain and therefore cannot be reliably applied to the forward region.

The parameter values obtained from the fits are displayed in Fig.~\ref{fig:fig6}. For the sake of a simpler notation, the symbol $N_0$ is introduced in this figure to represent $\left.dN/dy\right|_{y=0}$. The results suggests that the model provides a good description of the data in both centrality classes. The extracted effective temperature values are consistent with the measurements published by the PHENIX Collaboration~\cite{PHENIX:2003iij}. Moreover, this agreement between data and theory was achieved while ensuring that the values of $\lambda$ and $C_{\rm h}$ remain compatible with the $\alpha_{\rm h}$ values inferred from the analysis of the PHENIX non-prompt photon spectra. Correspondingly, the uncertainties of the parameter $C_{\rm h}$ were estimated via error propagation, based on the uncertainties of $\lambda$ and $\alpha_{\rm h}$.

All of this suggests that the model presented in the present manuscript can describe thermal radiation while also providing a reasonable representation of the experimental observations in the hadronic channel.

\section{Equation of State Predicted by the Model}\label{sec:6}

In this section, we examine the equation of state determined by the parameters obtained from the fits and those previously fixed, based on eqs.~\eqref{eq:eos_def_hadron} and ~\eqref{eq:eos_def_quark}, which arise from one of the possible conditions for the analytic solvability of the hydrodynamic equations.

Taking into account the properties of the lattice QCD equation of state, the parameters $T_{\rm tr}$ and $\alpha_{\rm q}$ were fixed during the fits to the non-prompt direct photon spectra. The former was set to 2, motivated by the fact that $C_{\rm q}$ takes a value 3 in the conformal limit, while $\lambda$ can be estimated to be approximately 1.3 based on previous analyses of experimental data. The present analysis also supports a value of $\lambda \approx 1.3$ in 200 GeV $Au+Au$ collisions, as shown in Section~\ref{sec:5}. From the fit to the non-prompt component of the PHENIX direct photon spectra, the value of the parameter $\alpha_{\rm h}$ was determined and subsequently imposed as a constraint in the fits to the PHOBOS $dN/d\eta$ data, allowing the extraction of $C_{\rm h}$ and $\lambda$. Since $\alpha_{\rm q}$ also depends on $\lambda$ according to eq.~\eqref{eq:alpha_i}, the value of $C_{\rm q}$ could be precisely determined within the model, resulting in a slightly higher value than predicted by the conformal limit. This discrepancy can be attributed to the present model, which predicts an initial temperature below 500 MeV, where the conformal limit is not yet established, and thus is not "felt" by the experimental data. We will see that this finding is also supported by lattice QCD simulations.

As indicated by eq.~\eqref{eq:eos_def_hadron}, the evaluation of $\kappa_{\rm h}(T)$ further necessitates the specification of the kinetic freeze-out temperature $T_{\rm fo}$ and the corresponding parameter $\kappa_{\rm fo} = \kappa_{\rm h}(T_{\rm fo})$. The range of $T_{\rm fo}$ values quoted in the literature is rather broad~\cite{PHENIX:2003iij,STAR:2008med,STAR:2017sal,Burward-Hoy:2002zye,STAR:2004qya,Sett:2015lja,Yang:2025gbh}. In central $Au+Au$ collisions at 200 GeV, STAR reports a value of approximately 90 MeV~\cite{STAR:2008med}, whereas PHENIX predicts 177 MeV in ref.~\cite{PHENIX:2003iij}. In addition, the centrality dependence of the kinetic freeze-out temperature remains unsettled. For instance, refs.~\cite{Sett:2015lja} and~\cite{Yang:2025gbh} suggest that $T_{\rm fo}$ can be lower in peripheral than in central collisions, whereas the PHENIX collaboration, as well as the STAR collaboration in refs.~\cite{Burward-Hoy:2002zye,STAR:2008med,STAR:2017sal}, arrive at the opposite conclusion.

\begin{figure*}[!h]
\subfloat[0–20\% Centrality Prediction]{\includegraphics[width=0.49\linewidth]{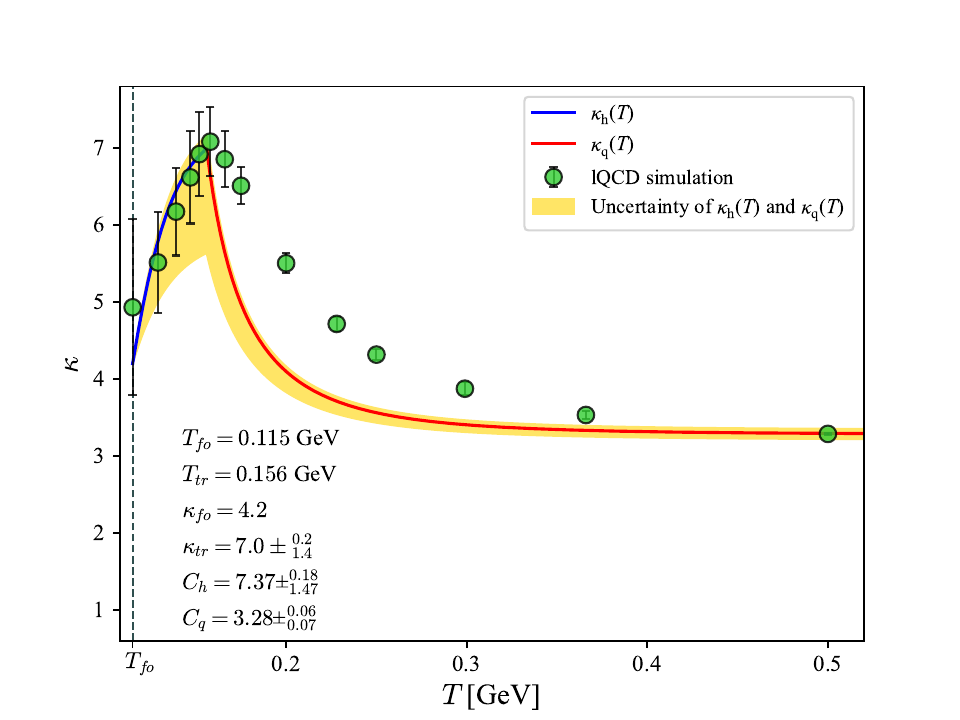}}
\hfil
\subfloat[20–40\% Centrality Prediction]{\includegraphics[width=0.49\linewidth]{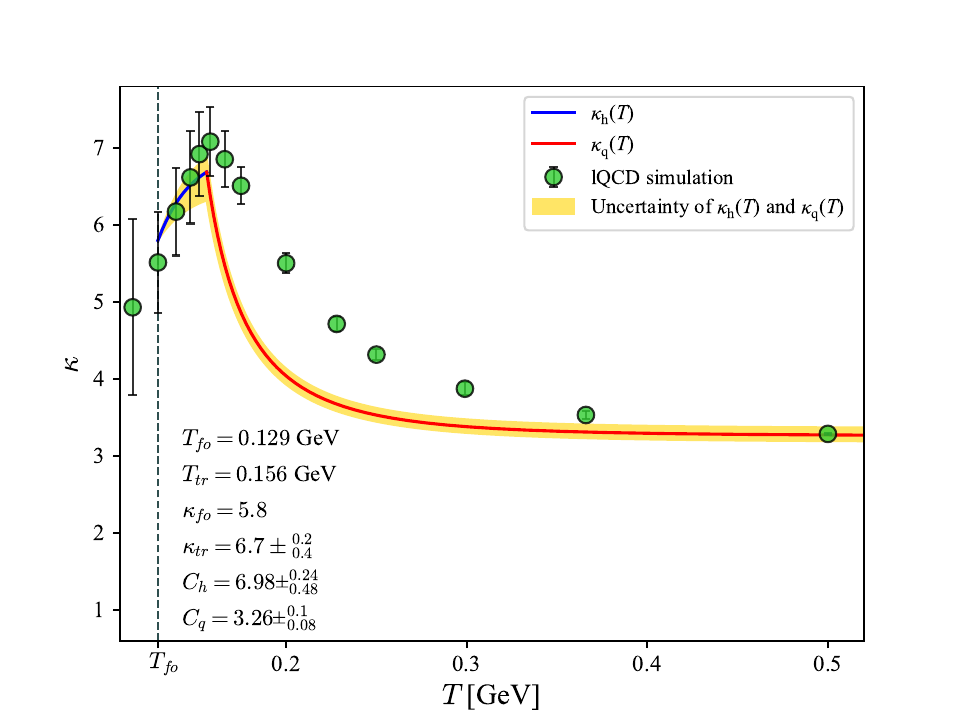}}
\caption{
The blue and red solid lines represent the outcome of my hydrodynamic model for the energy density to pressure ratio in the hadronic and QGP regimes, respectively, based on analyses carried out for the 0–20\% (left) and 20–40\% (right) centrality classes. The parameters of the red and blue curves were constrained both by fits to experimental data and by the lattice QCD equation of state. The theoretical uncertainty arising from the parameter errors is indicated by the yellow band. The vertical dashed line indicates the location of the kinetic freeze-out temperature. The green data points show the prediction of lattice QCD simulations~\cite{Borsanyi:2010cj}, including their associated uncertainties.}
\label{fig:fig7}
\end{figure*}

In this work, I adopted the approach that the kinetic freeze-out temperature $T_{\rm fo}$ is lower in more central collisions, while selecting intermediate values relative to the more extreme estimates reported in the literature. For simplicity, the values of $T_{\rm fo}$ and $\kappa_{\rm fo}$ were fixed in both centrality classes near specific points of lattice QCD simulations, serving as effective anchor points. The chosen values are indicated in Fig.~\ref{fig:fig7}. This figure illustrates the functions $\kappa_{\rm h}(T)$ and $\kappa_{\rm q}(T)$ implied by eqs.~\eqref{eq:eos_def_hadron} and~\eqref{eq:eos_def_quark} when using the parameters ($C_{\rm h}$ and $C_{\rm q}$) extracted from the analyses of the centrality classes 0–20\% and 20–40\%, together with those fixed by the lattice QCD equation of state ($T_{\rm tr}$, $\kappa_{\rm tr}$ and $\kappa_{\rm fo}$). The peak of the curve resulting from the combination of these two functions is characterized by the value of $\kappa_{\rm tr}$, determined through the matching condition $\kappa_{\rm h}(T_{\rm tr}) = \kappa_{\rm q}(T_{\rm tr})$. The uncertainty of the parameter $C_{\rm h}$ was derived from the errors of $\lambda$ and $\alpha_{\rm h}$, while the uncertainty of $C_{\rm q}$ originates solely from that of $\lambda$, since the value of $\alpha_{\rm q}$ was fixed during the fit to the non-prompt photon spectra. Since the PHOBOS collaboration provided $dN/d\eta$ data at $\sqrt{s_{NN}}=200$ GeV $Au+Au$ collisions only for central and mid-central classes, no fitted value of $\lambda$ is available for more peripheral collisions. Consequently, the parameters $C_{\rm h}$ and $C_{\rm q}$ cannot be determined reliably in those cases, and the equation of state was therefore constructed only for the 0–20\% and 20–40\% centrality intervals.

In Fig.~\ref{fig:fig7}, alongside the curves constrained by the hydrodynamic equations ($\kappa_{\rm h}(T)$ and $\kappa_{\rm q}(T)$), the lattice QCD simulations reported in ref.~\cite{Borsanyi:2010cj} are also shown. It is clearly seen that on the low-temperature side of the peak, $\kappa_{\rm h}(T)$ is consistent with the lattice QCD points, whereas on the high-temperature side only a qualitative similarity is observed, with precise agreement emerging only at $T_{\rm tr}$ and at sufficiently high temperatures. The function $\kappa_{\rm q}(T)$ remains a strictly monotonically decreasing function in the QGP region, consistent with both the data point that defines the peak structure and the high-temperature asymptotic limit of the lattice QCD results. These are precisely the properties that are relevant from the perspective of the hydrodynamic model. In the hydrodynamic solution on the QGP side, the effect of $\kappa_{\rm q}(T)$ enters through the parameter $C_{\rm q}$, i.e. through the high-temperature asymptotic behavior. Furthermore, for a proper matching between the hadronic and QGP sides, it is essential that the model can also reproduce the location of the peak — a feature that the present parametrization of the function $\kappa_{\rm tot}(T)$ successfully achieves.

Additionally, it should be noted that the aim was not to reproduce the lattice QCD equation of state per se, but rather to use it as a guide for the parametrization of the functions $\kappa_{\rm q}(T)$ and $\kappa_{\rm h}(T)$, and to assess whether, under such constraints, the model is capable of describing the PHENIX non-prompt direct photon spectra.

\begin{figure*}[!h]
\subfloat[0–20\% Centrality Prediction]{\includegraphics[width=0.49\linewidth]{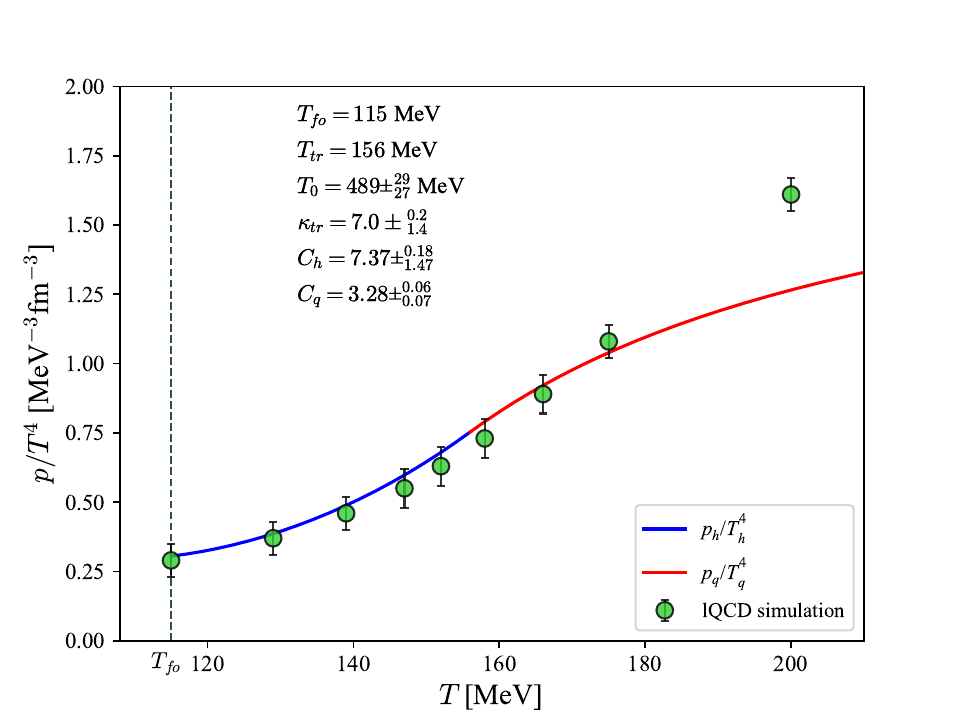}}
\hfil
\subfloat[20–40\% Centrality Prediction]{\includegraphics[width=0.49\linewidth]{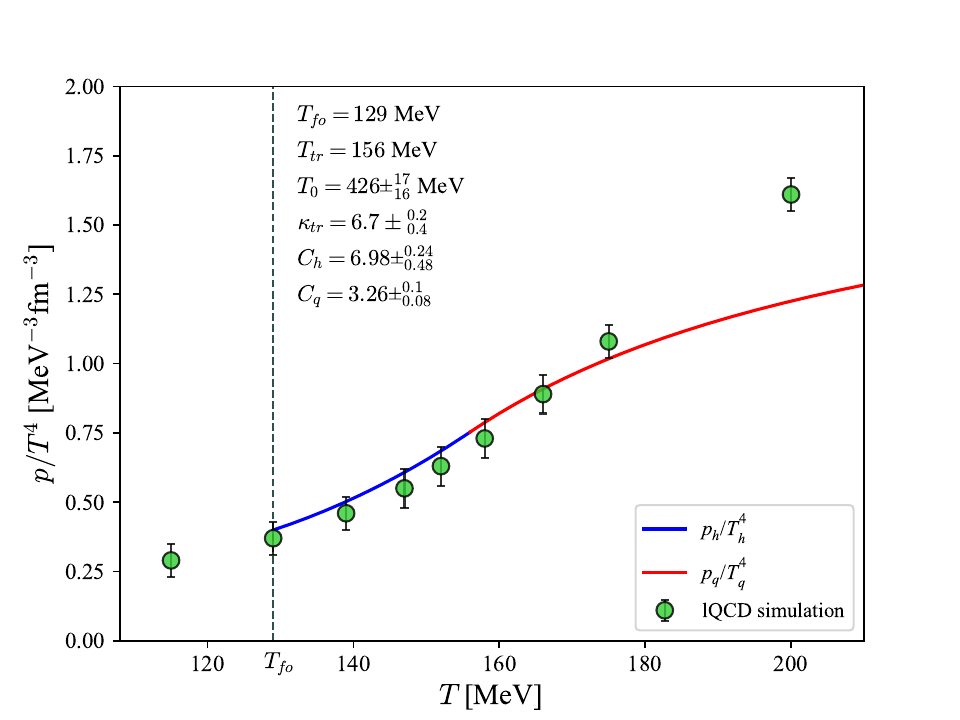}}
\caption{
The blue and red solid lines represent the hydrodynamic model predictions for the quantity $p/T^4$ in the hadronic and QGP regimes, respectively, based on analyses carried out for the 0–20\% (left) and 20–40\% (right) centrality classes. The parameters of the red and blue curves were constrained both by fits to experimental data and by the lattice QCD equation of state. The vertical dashed line indicates the location of the kinetic freeze-out temperature. The green data points show the prediction of lattice QCD simulations~\cite{Borsanyi:2010cj}, including their associated uncertainties.}
\label{fig:fig8}
\end{figure*}

It is important to emphasize that the functions $\kappa_{\rm h}(T)$ and $\kappa_{\rm q}(T)$ characterize the behavior of the ratio of energy density to pressure. Consequently, the consistency of the high-temperature limit of $\kappa_{\rm q}(T)$ with lattice QCD simulations does not necessarily imply that the same level of agreement holds for the high-temperature behavior of the pressure or the energy density themselves. Figure~\ref{fig:fig8} displays the quantity $p/T^4$ around the transition region, where the curve obtained from the hydrodynamic equations is consistent with the lattice QCD results, and also slightly beyond, where this agreement begins to break down. For temperatures above 200 MeV, the consistency deteriorates significantly, and at higher temperatures only the overall magnitude is reproduced.

The blue–red curve derived from the hydrodynamic framework is, however, highly sensitive to the parameter $C_{\rm q}$, which takes the values 3.28 (for centrality 0-20\%) and 3.26 (for centrality 20-40\%). If instead $C_{\rm q}$ were chosen to be around 4, it would enhance the consistency of the outcomes from my model for $p/T^4$ with lattice QCD simulations in the region $T < 500$ MeV, although at the expense of losing the consistency of $\kappa_{\rm q}(T)$ with the conformal limit at high temperatures. Although such a parameterization would lead to slightly better agreement between $\kappa_{\rm q}(T)$ and the lattice QCD calculations, it would typically reproduce the asymptotic limit of $\varepsilon/p$ at high temperatures with considerably less accuracy. While these forms may better match the lattice QCD pressure values, but for the purposes of the present analytic model the temperature evolution is the key feature of interest. In the differential equation that governs the temperature (eq.~\eqref{eq:energy-cons}), the effect of the pressure enters the dynamics solely through the function 
$\kappa(T)=\varepsilon/p$, without an additional explicit dependence on $p$.

A detailed investigation of this issue would require a new fitting strategy for the non-prompt direct photon spectra, since in the fits presented in Section~\ref{sec:4} the parameter $\alpha_{\rm q}$ was fixed precisely to keep $C_{\rm q}$ close to 3, thereby ensuring the correct high-temperature limit of $\kappa_{\rm q}(T)$. A parameterization in which the value of $C_{\rm q}$ is increased to approximately 4 would predict a higher initial temperature; however, the difference would remain below 10\% and within the statistical uncertainties. For this reason, I have opted for the parameterizations illustrated in Fig.~\ref{fig:fig7}, which provide a more reliable description of the high-temperature asymptotic behavior, consistent with lattice QCD expectations.

A detailed analysis of this issue is planned for a future publication, as it goes beyond the scope of the present manuscript.

\section{Discussion}\label{sec:7}
Based on the results presented in Section~\ref{sec:4}, it can be concluded that the model provides a qualitative description of the non-prompt direct photon spectra measured in PHENIX $Au+Au$ collisions at $\sqrt{s_{\rm NN}} = 200$ GeV. This agreement is achieved without accounting for the effects of viscosity, suggesting that viscous effects do not significantly influence the shape of the thermal radiation spectrum. Moreover, since transverse dynamics are not included in the framework, the agreement with the experimental data — as illustrated by Figures~\ref{fig:fig1} through~\ref{fig:fig4} — also suggests that the thermal radiation is dominated by the contribution of QGP, where the effect of radial flow is comparatively less pronounced. Nevertheless, incorporating such effects would naturally provide a more refined estimate of the initial temperature (the effect of radial flow is expected to reduce the value of $T_0$). Moreover, by extending the dynamics of the model to the transverse directions, it would also become possible to compute the $v_2$ elliptic flow coefficient within the same framework. However, due to the analytic nature of the model, incorporating such effects is a highly non-trivial task and is therefore beyond the scope of this manuscript.

In refs.~\cite{PHENIX:2022rsx,Orosz:2025cey,Esha:2022tvk}, the PHENIX collaboration reported, among other observations, that they fit the dataset represented with green markers in Figures~\ref{fig:fig1}–\ref{fig:fig4} using simple exponential functions in order to extract the inverse slope parameter ($T_{\rm eff}$), commonly referred to as the effective temperature of the spectrum. This parameter does not directly represent any temperature of the medium, it rather corresponds to a temperature averaged over space and time (complemented by the effect of radial flow, which is less significant than the temperature in the high-$p_{\rm T}$ region~\cite{Kiyomichi:2005va,ALICE:2025iud}). This interpretation is consistent with the finding that the initial temperature values ($T_0$) obtained in the present analysis exceed the effective temperatures reported by PHENIX in refs.~\cite{PHENIX:2022rsx,Orosz:2025cey,Esha:2022tvk}. They demonstrated that the extracted values of $T_{\rm eff}$ depend on the chosen fit range, thus the behavior of the measured data cannot be adequately described by a simple exponential spectrum. However, they have not found a significant deviation from the hypothesis of a constant $T_{\rm eff}$ across centrality classes. Figure~\ref{fig:fig5} of the present study examines the centrality dependence of the initial temperature $T_0$, revealing that a centrality-independent $T_0$ remains consistent with the data (owing to the more significant statistical and systematic uncertainties) only if the contribution of hadronic sources is taken into account in the model. When this hadronic component is neglected, a centrality-independent initial temperature becomes incompatible with the experimental observations.

As mentioned in an earlier section of this manuscript, the present model, which relies on several simplifying assumptions, differs from approaches based on relativistic kinetic theory. Certain aspects of the discrepancy between the two frameworks can be estimated using the method described in refs.~\cite{Csanad:2011jq, Lokos:2024yjm}. According to ref.~\cite{Lokos:2024yjm}, accounting for this difference introduces an uncertainty of less than 2\% in the initial temperature extracted, which remains smaller than the uncertainty arising from statistical errors. Based on these considerations, this source of systematic uncertainty is deemed negligible in the present analysis. However, further discrepancies between the two approaches may also be present (such as the logarithmic prefactor discussed in ref.~\cite{Shen:2013vja}) the systematic implications of which would merit careful investigation in future work. Nonetheless, a supporting example can be found in ref.~\cite{Turbide:2003si}, where the authors investigated hadronic thermal radiation by calculating the rates of various microscopic processes. The results of that study are consistent with those presented here, in the sense that both models predict the hadronic component to exceed the QGP contribution in the $p_{\rm T}<1$ GeV region of $Au+Au$ collisions at 200 GeV. Similarly to the previous case, ref.~\cite{Turbide:2007mi} provides another supporting example, where for 200 GeV $Au+Au$ collisions at 0–20\% centrality, it was likewise predicted that around $p_{\rm T} \approx 1$ GeV the hadronic component of the thermal emission outshines the thermal QGP contribution.

In ref.~\cite{Kasza:2023rpx}, a simplified, single-component version of the present analytic model was fitted to the 0–20\% centrality data set, which is also used in the current manuscript. That earlier model predicted an exceptionally high initial temperature, accompanied by large statistical and systematic uncertainties, yielding $T_0 = 716\pm^{327}_{173}(\textnormal{stat.})\pm^{32}_{238}(\textnormal{syst.})$ MeV. In the present work, by neglecting the hadronic contribution during the fit, the applied framework effectively reduces to the single-component model of ref.~\cite{Kasza:2023rpx}, but with one crucial refinement: instead of assuming a constant $\kappa$, the parameter $C_{\rm q}$ is introduced. This parameter can be interpreted within a physically motivated context, allowing its value to be fixed owing to the extended two-component formulation. Such constraints were not available in the single-component analysis of ref.~\cite{Kasza:2023rpx}; consequently, neither $\kappa$ nor the derived parameter ($\alpha = 2\kappa/\lambda-3$) could be fixed or even meaningfully bounded. As a result, $\alpha$ — corresponding to $\alpha_{\rm q}$ in the present formulation — was treated as a free parameter in ref.~\cite{Kasza:2023rpx}, leading to a higher central value for $T_0$ and significantly larger uncertainties.

In refs.~\cite{Csanad:2011jq,Csanad:2011jr}, an analytical model was similarly applied to describe the direct photon yield in PHENIX $Au+Au$ collisions at $\sqrt{s_{\rm NN}} = 200$ GeV. The model was based on a 1+3 dimensional analytical hydrodynamic solution featuring a locally accelerationless velocity field, and predicted a lower limit of $T_0 > 507 \pm 12$ MeV for the initial temperature of the medium produced in 0–20\% centrality collisions. This result is consistent within uncertainties with my own finding that includes the hadronic contribution. In contrast, neglecting the hadronic component reduces the estimated initial temperature to such an extent that it becomes incompatible with the aforementioned lower bound. However, in refs.~\cite{Csanad:2011jq,Csanad:2011jr}, the comparison was made to the complete direct photon spectrum~\cite{PHENIX:2008uif}, since experimental data separating the non-prompt component had not yet become available at the time. Consequently, the fit was restricted to the low-$p_{\rm T}$ region of $p_{\rm T} < 3.5$ GeV, which already exceeds the domain dominated by the thermalized medium. The final data point included in their fit is no longer dominated by thermal photons and tends to pull up the tail of the fitted curve, ultimately leading to the prediction of a higher initial temperature. Although the model presented in refs.~\cite{Csanad:2011jq,Csanad:2011jr} assumes a boost-invariant velocity field, it remains highly valuable, as it is based on a 1+3 dimensional solution and thus incorporates information on the transverse dynamics as well. Consequently, it is capable of describing the elliptic flow parameter $v_2$ of direct photons. Therefore, it would be worthwhile to revisit the analysis detailed in refs.~\cite{Csanad:2011jq,Csanad:2011jr}, this time using the more recent experimental data in which the non-prompt component has been separated in the direct photon spectrum, which also serve as the basis for the model comparisons presented in this work.

\section{Summary}\label{sec:8}
In this manuscript, I have presented a novel analytic model aimed at describing the transverse momentum ($p_{\rm T}$) spectrum of thermal radiation originating from heavy-ion collisions. The model is based on a relativistic solution of perfect fluid hydrodynamics which, although limited to 1+1 dimensions, is characterized by a locally accelerating velocity field. Due to the absence of transverse dynamics, the model does not account for elliptic flow; nevertheless, it successfully describes PHENIX data from $\sqrt{s_{\rm NN}} = 200$ GeV $Au+Au$ collisions across four different centrality classes — even in the more peripheral classes, where the applicability of hydrodynamics is less justified. However, owing to the number of simplifying assumptions inherent in the model, the comparisons presented in this paper are primarily indicative, illustrating overall tendencies and trends, while also providing a useful reference point for more sophisticated numerical calculations. A more realistic equation of state, together with a fully three-dimensional evolution, would be required to support quantitatively robust conclusions.

A key novel feature of the model lies in its explicit treatment of the quark–hadron transition, allowing the thermal emission to be decomposed into two distinct components: one from a higher-temperature (QGP) phase and one from a lower-temperature (hadronic) phase. I performed two sets of fits to the PHENIX data, motivated by the observation that the temperature $T_{\rm f}$ (where the thermal photon emission has already come to a halt) tends to converge to the transition temperature $T_{\rm tr}$. In one approach, I simply neglected the hadronic contribution by treating it as negligible. In the second approach, $T_{\rm f}$ was fixed to a value close to $T_{\rm tr}$, thus allowing for the inclusion of the hadronic component of thermal photons. I found that in the two most central classes — where the applicability of hydrodynamics is better justified — the two-component model provided an improved description of the data. When the hadronic contribution was included, the initial temperature values extracted $T_0$ were higher, and their statistical and systematic uncertainties were sufficiently large to preclude any definitive conclusion regarding the centrality-independence of $T_0$. In contrast, when the hadronic component was neglected, the uncertainties decreased and the structure of the centrality dependence of $T_0$ became significantly more pronounced.

Although the present manuscript focuses on describing thermal radiation, the new model was also tested in the hadronic channel. I found that it is capable of describing the PHOBOS pseudorapidity density measurements with good confidence, demonstrating that the model is simultaneously applicable to both the photonic and hadronic channels.

Building on the parameter values obtained from fits to both photonic and hadronic observables, and guided by lattice QCD simulation results, I was able to parametrize the equation of state whose functional form is constrained by the conditions required for the solvability of the conservation equation of energy. The resulting equation of state agrees with lattice QCD calculations within uncertainties at low temperatures ($T \lesssim T_{\rm tr}$), whereas at higher temperatures, agreement is merely qualitative and becomes apparent only in the ratio of energy density to pressure.

References~\cite{Csanad:2011jq,Csanad:2011jr,Lokos:2024yjm} discuss a model derived from a 1+3 dimensional, boost-invariant hydrodynamic solution. A major advantage of this approach lies in its ability to compute both the elliptic flow parameter ($v_2$) and the $p_{\rm T}$ spectrum of thermal photons within a unified and tractable framework — an essential feature for addressing the so-called "direct photon puzzle". However, the absence of acceleration in the velocity field requires a larger spatial eccentricity and a longer-lived medium to match the observed $v_2$ data. This suggests that incorporating acceleration into the velocity field may alleviate the need for such extreme conditions. Consequently, there is a clear motivation to pursue the 1+3 dimensional generalization of the hydrodynamic solution introduced in this work. This is particularly important given that, beyond the elliptic flow and the $p_{\rm T}$ spectrum, there are currently few experimental observables available for testing models of thermal photon production. One potential candidate is the nuclear modification factor ($R_{\rm AA}$) of direct photons; however, within the present model this quantity is identically unity — as a direct consequence of the functional form assumed for the transparency $H(\tau)$ — which is not inconsistent with existing data~\cite{PHENIX:2012jbv}. Another potential observable could be the HBT radii of direct photons, though to the best of my knowledge, no such experimental results are currently available.

The model presented in this manuscript does not include viscous effects, yet it remains consistent with experimental data. This naturally raises the question of whether including viscosity would have a significant impact on the description of thermal photon observables. To address this question within an analytic framework, one requires an analytic solution of relativistic viscous hydrodynamics. This highlights the need for a viscous generalization of the solutions introduced in Section~\ref{sec:solution}. A further exciting possibility lies in the extension of the model discussed in refs.~\cite{Csanad:2011jq,Csanad:2011jr,Lokos:2024yjm}: the hydrodynamic solution employed in those studies already possesses known viscous generalizations~\cite{Csanad:2019fto,Csanad:2019lcl,Csorgo:2020iug}. This opens a pathway toward studying thermal photon emission within a completely analytic framework that incorporates viscous corrections.

\section*{Acknowledgements}
This research was funded by NKFIH grants K-138277 and K-147557, and the KKP-2024 Research Excellence Programme of MATE, Hungary.

\let\doi\relax
\printbibliography

\end{document}